\documentclass[11pt,a4paper]{article}
\pdfoutput=1
\usepackage{jheppub}
\usepackage[utf8]{inputenc}
\usepackage{soul}
\usepackage{ulem}
\usepackage{graphicx}
\usepackage{comment}
\usepackage{bm,array}
\usepackage{graphics}
\usepackage{epsf}
\usepackage{color}
\usepackage{amsmath}
\usepackage{amssymb}
\usepackage{latexsym}
\usepackage{slashed}
\usepackage{float}
\usepackage{cases}
\usepackage{multirow}
\def\ba{\begin{array}}
\def\ea{\end{array}}
\def\l{\lambda}
\def\k{\kappa}
\def\mueff{\mu_{\mathrm{eff}}}
\def\sQ3{\tilde{Q}_3}
\def\sU3{\tilde{U}_3}
\def\sD3{\tilde{D}_3}
\def\stleft{\tilde{t}_L}

\def\sbleft{\tilde{b}_L}
\def\sbright{\tilde{b}_R}
\def\stone{\tilde{t}_1}
\def\sttwo{\tilde{t}_2}
\def\stonetwo{\tilde{t}_{1,2}}
\def\sbone{\tilde{b}_1}
\def\sbtwo{\tilde{b}_2}
\def\sbonetwo{\tilde{b}_{1,2}}
\def\ntrli{\chi_i^0}

\def\ntrlone{\chi_1^0}
\def\ntrltwo{\chi_2^0}
\def\ntrlthree{\chi_3^0}
\def\ntrlfour{\chi_4^0}
\def\ntrlfive{\chi_5^0}

\def\ntrltwothree{\chi_{2,3}^0}

\def\ntrlonetwothree{\chi_{1,2,3}^0}

\def\charonep{\chi_1^+}
\def\charonem{\chi_1^-}

\def\charonemp{\chi_1^\mp}
\def\msQ3{m_{\tilde{Q}_3}}
\def\msU3{m_{\tilde{U}_3}}
\def\msD3{m_{\tilde{D}_3}}
\def\mstleft{m_{\tilde{t}_L}}
\def\mstright{m_{\tilde{t}_R}}
\def\msbleft{m_{\tilde{b}_L}}
\def\msbright{m_{\tilde{b}_R}}
\def\mstone{m_{\tilde{t}_1}}
\def\msttwo{m_{\tilde{t}_2}}
\def\msbone{m_{\tilde{b}_1}}
\def\msbtwo{m_{\tilde{b}_2}}

\def\mntrltwo{m_{\tilde{\chi}_2^0}}

\def\mcharonem{m_{\tilde{\chi}_1^-}}
\def\mcharonepm{m_{\tilde{\chi}_1^\pm}}
\def\mchartwopm{m_{\tilde{\chi}_2^\pm}}
\def\costhetab{\cos\theta_{\tilde{b}}}
\newcommand{\fbinv}{\text{fb}$^{-1}$}
\def\etmiss{\slashed{E}_T}
\def\mttwo{m_{T2}}
\newcommand{\beq}{\begin{equation}}
\newcommand{\eeq}{\end{equation}}
\newcommand{\bea}{\begin{eqnarray}}
\newcommand{\eea}{\end{eqnarray}}
\title{Sbottoms of natural NMSSM at the LHC}
\author[a, b]{Jyotiranjan Beuria}
\author[c]{Arindam Chatterjee}
\author[a]{AseshKrishna Datta}

\affiliation[a]{Harish-Chandra Research Institute, Allahabad 211019, India}
\affiliation[b]{Regional Centre for Accelerator-based Particle Physics \\
                Harish-Chandra Research Institute, Allahabad 211019, India} 
\affiliation[c]{Physics and Applied Mathematics Unit, Indian Statistical Institute,\\
                203 B.T. Road, Kolkata 700108, India}

\emailAdd{jyotiranjan@hri.res.in, arin\_t@isical.ac.in, asesh@hri.res.in} 
\preprint{HRI-P-16-03-001 \\ 
\vspace*{-0.8cm}
\begin{flushright}
RECAPP-HRI-2016-004
\end{flushright}
}
\abstract{Search for the bottom squarks (sbottoms) at the Large Hadron Collider
(LHC) has recently assumed a heightened focus in the hunt for Supersymmetry
(SUSY). The popular framework of the Next-to-Minimal Supersymmetric Standard
Model (NMSSM) could conceive a naturally light sbottom which can easily be
consistent with available constraints from the experiments at the LHC.
Phenomenology of such sbottoms could in principle be as striking as that for a
light top squark (stop) thanks to a rather nontrivial neutralino sector (with
appreciable mixing among the neutral higgsinos and the singlino) that the
scenario gives rise to.  Nonetheless, finding such sbottoms would require a
moderately large volume of data ($\sim 300$ \fbinv) at the 13 TeV run of the
LHC. A multi-channel analysis establishing a generic depletion of events in the
usual $2b$-$jets +\etmiss$ final state while registering, in conjunction,
characteristically significant rates in various multi-lepton final states
accompanied by $b$-$jets$ might point not only to the presence of light
sbottom(s) but could also shed crucial light on their compositions and the
(singlino) nature of the lightest SUSY particle (LSP).}
\begin{document}
\maketitle
%
\section{Introduction}
\label{sec:intro}
It is well known that squarks from the third generation, in particular, the
top squarks (stops), are capable of providing a `natural' supersymmetric
(SUSY) solution to the notorious hierarchy problem that cripples the scalar
(Higgs) sector of the Standard Model (SM) if they are relatively light
($\lesssim 1 \; \mathrm{TeV} $). Over the years, this has motivated the search 
for sub-TeV stops at the collider experiments and recently, at the Large
Hadron Collider (LHC), hunt for them has expectedly taken the centre stage.
The latest analyses of the  data from the 7 TeV and 8 TeV runs of the LHC
(collectively termed as the LHC Run-I in this work) 
\cite{Khachatryan:2015wza,Aad:2015pfx} have resulted in stringent lower bounds
on the stop mass, under varied assumptions over how it decays. The most
conservative of these bounds says that the lighter stop should be heavier
than $\sim 700$ GeV \cite{Aad:2015pfx}. A reasonably performing 13 TeV run of 
the LHC (LHC-13), currently underway, could quickly and substantially improve 
on these bounds thus pushing them around a TeV.

On the other hand, over the past few years, there has been a growing interest
in the phenomenology of the bottom squarks (sbottoms)
\cite{Li:2010zv, Datta:2011ef,Lee:2012sy,Alvarez:2012wf,Bi:2012jv,Ghosh:2012wb,
Chakraborty:2013moa,Belanger:2013gha,Batell:2013psa, Huang:2014oza,
Kobakhidze:2015dra,Dutta:2015hra,Han:2015tua},
%
the other squarks from the third generation. Also, in recent years, 
implications of the LHC results for the squarks from the third generation
in a `natural' SUSY setup {\cite{Barbieri:1987fn}} are studied intensively
\cite{Cheng:2013fma,Han:2013kga, Zheng:2014dea, Grothaus:2015yha}.
This has a solid motivation. Although the sbottom sector is not broadly as 
instrumental as the stop sector for rendering SUSY `natural' (due to its 
relatively much smaller coupling to the Higgs boson), sbottoms are rather close 
cousins of the stops. In particular, the left-handed (L-type) stop and the sbottom
($\stleft$ and $\sbleft$) live in the same weak doublet and enjoy an approximate
custodial symmetry \cite{Lee:2012sy}. Quantitatively, its breaking is restricted
by electroweak precision data via the well-known Peskin-Takeuchi $STU$ (oblique)
parameters \cite{Peskin:1990zt}.  On the theoretical side, the soft masses for
these L-type states are $SU(2)_{L}$-invariant and hence equal. However, a modest
splitting occurs due to $SU(2)_{L}$-breaking $D$-term (which is a function of
$\tan\beta ={v_u \over v_d}$, the ratio of vacuum expectation values of the 
neutral components of the two Higgs doublets) and because of the presence of an 
$F$-term which goes as the corresponding quark (top or bottom) mass.
This implies that at least one squark state from each flavor (which are
dominantly L-type) would have comparable masses. Only a rather large mixing in
the stop sector could obscure this phenomenon under certain circumstances.
On the other hand, the masses of the right-handed (R-type) stop and the sbottom
($\mstright$ and $\msbright$) are neither related by any symmetry nor do they
bear any relation to their corresponding L-type partners. Thus, they can assume
any value allowed by applicable searches at the colliders. Hence if the lighter
stop is accessible to the LHC experiments and happens to be mostly L-type, the
same should be true for a sbottom which is also dominantly L-type. This
substantiates the increased interest in the search for sbottoms at the LHC.
{On the other hand, a relatively light R-type sbottom can be present while the other
       sbottom and both stops are relatively heavy\footnote{ {\footnotesize It may, however, be noted that
      presence of relatively heavy stops (and/or a gluino) may not be an immediate 
      threat to naturalness as revealed by certain recent parameterizations 
      in the MSSM context
      that refer only to 
      weak scale values of the parameters as long as $|\mu|$ ($|\mu_{eff}|$, in NMSSM) 
      is small enough
      which is essential to the scenario
      presented in this work \cite{Baer:2012up, Baer:2013ava, Mustafayev:2014lqa, Baer:2015rja}.} }.}

It is somewhat interesting to note that various possible decay modes of the
sbottom have recently been explored rigorously for the first time by the
experimental searches at the LHC 
\cite{Chatrchyan:2013fea,Aad:2014qaa,Aad:2014pda,Aad:2015pfx,Khachatryan:2015wza}.
These result in lower bounds on $\msbone$ roughly ranging between 500-750 GeV.
Subsequently, these findings are exploited by the phenomenological works
providing further insight \cite{Han:2015tua,Ghosh:2012wb}. However, these
studies are heavily based on the popular framework of the Minimal SUSY extension
of the Standard Model (MSSM). In the present work, we look into the phenomenology
of possibly light sbottom(s) in the framework of the next-to-MSSM (NMSSM)
\cite{Ellwanger:2009dp}, a scenario which has assumed a renewed relevance in view
of the recent Higgs results from the LHC.

Issues with sbottom search vis-a-vis stop search have already been discussed in
the literature \cite{Bai:2012gs}. For example, search for stop squarks may suffer
from large background from the SM top quark production when the decay
$\stone \to t \ntrlone$ is dominant. This is not the case with sbottom search in
one of its preferred decay modes, i.e., $\sbone \to b \ntrlone$. On the other
hand, depending on how they decay, stop and sbottom could lead to the same final
states thus making it somewhat difficult to extract unambiguous information.
Combining analyses for the sbottom and stop to find an optimal sensitivity to a
given region in the parameter space parameter has recently been advocated in
reference \cite{Han:2015tua}.

The popular framework of the NMSSM offers new possibilities in this context.
First, the scope of having a richer electroweak gaugino sector (a mixture of a
singlino, higgsinos and gauginos) with relatively light states could lead to
interesting phenomenology of the stop and the sbottom through their decays to
these states. Secondly, the NMSSM can easily accommodate relatively light stops
without being in conflict with the observations in the Higgs sector
\cite{Beuria:2015mta}. If such a light stop happens to be dominantly L-type in
nature, there is going to be an L-type sbottom close-by in mass to the light
stop. These two issues render the NMSSM to be a `natural' setup to study the
sbottom sector. Interesting aspects of such a scenario and their
phenomenology have recently been pointed out mainly in the context of (or in
terms of) relatively light stops accompanied by a singlino-like neutralino LSP
while the lighter chargino and a pair of immediately heavier neutralinos
are higgsino-like \cite{Beuria:2015mta, Kim:2015dpa}.

In the present work, we focus on relatively light sbottom(s) with varied
compositions whose phenomenology might turn out to be more tractable than that
of the stops. This is since the extent of mixing that could be present in the
sbottom sector is much smaller than that possible in the stop sector. We adopt
a setup inspired by naturalness and hence with low values of the effective $\mu$
parameter ($\mueff$) {\cite{Ellwanger:2011mu}}. This exploits the
original scope in the NMSSM to dynamically generate such a low value of $\mueff$
via a suitable choice of the vacuum expectation value ({\it vev}) $v_S$ of the
singlet scalar field, a defining feature of such a scenario. As we shall see,
this opens up further possibilities of not only having a singlino-like lightest
SUSY particle (LSP) but also nontrivial mixings among the singlino and the
higgsinos. Impact of the singlino on the NMSSM phenomenology under specific
circumstances have been discussed in the literature
\cite{Das:2012rr, Ellwanger:2013rsa, Kim:2014noa, Ellwanger:2014hia,
Cao:2014efa, Potter:2015wsa}.
Here, we investigate how the mixed singlino-higgsino LSP would alter the decay
pattern of the sbottom in characteristic ways. In particular, for a mostly L-type
sbottom, its decay to higgsino-like lighter chargino is driven by the top Yukawa
coupling and hence could dominate, in particular, at low $\tan\beta$. In
contrast, decays of a dominantly R-type light sbottom to lighter higgsinos-like
states (both chargino and neutralinos) are all governed by the bottom Yukawa
coupling and thus, could become comparable.

Light stops when accompanied by a modest value of $\mueff$ could further ensure
a lower fine-tuning in the setup. These together add another perspective to the
current issue. A stop lighter than what is achievable in the MSSM (being
compatible with the observed Higgs mass) can be obtained for larger values of
the $\lambda$ parameter in the NMSSM. Note that $\lambda$ couples the singlet
state to the doublet Higgs states in the NMSSM superpotential. However, given
the relationship $\mueff = \lambda v_S$, a smaller value of $\mueff$ would then
require a relatively small $v_S$. On the other hand, avoiding appearance of
Landau pole below the unification (GUT) scale restricts $\kappa$ to smaller
values when $\lambda$ is large. As we will see later, a small $\mueff$ combined
with smaller $\kappa$ but a large $\lambda$, could induce a significant mixing
among the higgsinos and the singlino before reaching a point when the LSP becomes
singlino-like with $m_{LSP} \approx m_{\tilde{S}} = 2 \kappa v_S$.

The paper is organized as follows. In section \ref{sec:setup} we outline the
theoretical setup for the $Z_3$-invariant NMSSM by briefly discussing the
sbottom (stop) sector, the Higgs sector, the sectors involving the charginos and
the neutralinos and the relevant interactions among these states which are
germane to the present work. In section \ref{sec:sbot-lhc1} we look into the
status of sbottom search at the end of the LHC Run-I when interpreted in the NMSSM
framework. We take a critical look into the cases with sbottoms(s) having varied
compositions and their phenomenological ramifications. Section
\ref{section:benchmark-lhc13} is devoted to identifying a few benchmark scenarios
based on the findings of section \ref{sec:sbot-lhc1} followed by the presentation
of their detailed simulation at the 13 TeV run of the LHC. In section
\ref{sec:conclusions} we conclude.
%
\section{The theoretical setup}
\label{sec:setup}
%
The present work relies on a popular NMSSM framework known as the $Z_3$-invariant
NMSSM. The superpotential is given by
\beq
\mathcal{W}=\mathcal{W_{MSSM}}|_{\mu=0}+ \lambda \hat{S} \hat{H}_u.\hat{H}_d
  + {\kappa \over 3} \hat{S}^3.
\label{eq:superpot}
\eeq
Here, $\cal{W_{MSSM}}$ is the MSSM superpotential, $\hat{H}_{u}$ and $\hat{H}_{d}$
stand for the doublet Higgs superfields while the gauge singlet superfield,
characterizing the NMSSM, is denoted by $\hat{S}$. The corresponding
soft-supersymmetry breaking terms are given by
\beq
-\mathcal{L}_{\rm soft}= -\mathcal{L^{\rm MSSM}_{\rm soft}}|_{B\mu=0}+ m_{S}^2 |S|^2 +
\lambda A_{\lambda} S H_u.H_d
+ \frac{1}{3} \kappa A_{\kappa} S^3 + {\rm h.c.} +.. \,
\label{eq:ssbr}
\eeq
where  $\mathcal{L}^{\rm MSSM}_{\rm soft}$ represents the soft SUSY-breaking
terms in the MSSM, $A_{\lambda}$ and $A_{\kappa}$ are the new trilinear soft
SUSY-breaking terms (with dimensions of mass) that are present in the NMSSM while
$m_S^2$ is the soft SUSY-breaking mass term for the singlet scalar $S$ of the
NMSSM.

In the rest of this section we briefly discuss the salient features of the
sbottom and the stop sectors, the Higgs sector and the sector involving the
neutralinos and the charginos (the electroweakinos). The stop and the Higgs
sector play crucial roles in motivating and defining the overall scenario that
we work in while, in our `simplified' approach, except for some involved
situations, these are only the sbottoms and the electroweakinos that directly
shape up the interesting phenomenology pertaining to the sbottoms in the NMSSM.
%
\subsection{The sbottom and the stop sectors}
\label{subsec:sbot-stop}
%
At the tree level, the mass-squared matrices for the sbottom and the stop squarks
in the NMSSM are similar to the corresponding ones in the MSSM with $\mu$ in the
off-diagonal terms being replaced by $\mueff$ ($=\lambda v_S$). These are given,
at the tree level, by the following $2 \times 2$ matrices in the bases
$\{\msbleft, \msbright\}$ and $\{\mstleft, \mstright\}$, respectively;
\beq\label{msbottom}
{\cal M}_{\tilde{b}}=\left(\ba{cc} m_{\tilde{Q}_3}^2 + y_b^2 v_d^2 +
 (v_u^2-v_d^2)\left( \frac{g_1^2}{12}+\frac{g_2^2}{4}\right)
& y_b  (A_b v_d-\mueff v_u) \\
y_b (A_b v_d-\mueff v_u) & 
 m_{\tilde{D}_3}^2 + y_b^2 v_d^2 +(v_u^2-v_d^2)\frac{g_1^2}{6}
\ea \right)\;  
\eeq  \\
and
\beq\label{mstop} 
{\cal M}_{\tilde{t}}= \left(\ba{cc}  m_{\tilde{Q}_3}^2 + y_t^2 v_u^2 +
 (v_u^2-v_d^2)\left(\frac{g_1^2}{12}-\frac{g_2^2}{4}\right)
& y_t (A_t v_u - \mueff v_d)  \\
 y_t (A_t v_u - \mueff v_d) & 
m_{\tilde{U}_3}^2 + y_t^2 v_u^2-(v_u^2-v_d^2)\frac{g_1^2}{3}
\ea \right).
\eeq
\vskip 10pt
\noindent
In the above two equations $m_{\tilde{Q}_3}$, $m_{\tilde{D}_3}$ and
$m_{\tilde{U}_3}$ denote the soft SUSY breaking mass terms for the $SU(2)$
doublet (L-type) and the singlet (R-type) sbottom and stop squarks, respectively.
$y_{b,t}$ are the respective Yukawa couplings while $A_{b,t}$-s are the
respective trilinear soft SUSY breaking terms. $v_d$ and $v_u$ stand for the
{\it vev}'s of the CP-even down- and up-type neutral Higgs bosons, $H_d^0$ and
$H_u^0$, respectively. $g_1$ and $g_2$ denote the gauge couplings for the
$U(1)_Y$ and $SU(2)_L$ gauge groups, respectively. We also define a generic
rotation matrix $\mathcal{R}_{\tilde{f}}$ that diagonalizes the sbottom and stop mass
matrices as follows:
\beq
 {\cal R}_{\tilde{f}} = \left( 
                     \begin{array}{rr}
                       \cos\theta_{\tilde{f}} & \: \sin\theta_{\tilde{f}} \\
                      -\sin\theta_{\tilde{f}} & \: \cos\theta_{\tilde{f}}
                     \end{array}
              \right)
\eeq
where $\tilde{f}$ stands for $\tilde{b}$ or $\tilde{t}$, respectively and
$\theta_{\tilde{f}}$, at the tree level, is still given by its standard MSSM
expression
\beq 
\sin 2\theta_{\tilde{f}} = {2m_f X_f \over {m_{\tilde{f}_2}^2 - m_{\tilde{f}_1}^2}}
\quad, \qquad X_f =A_{\tilde{f}} -\mueff \; r \quad ,
\eeq
where $r= \tan\beta \, (\cot\beta)$ for the sbottom (stop) sector, $m_f$ is the
corresponding fermion mass while $m_{\tilde{f}_{1,2}}$ stand for the masses of
the lighter and the heavier sbottom/stop eigenstates, respectively. In this
convention, $\theta_{\tilde{f}}=0 ({\pi \over 2})$ corresponds to the unmixed
L-type (R-type) state to be the lightest mass eigenstate.

Clearly, if $m_{\tilde{Q}_3}$ is relatively small, one could expect a light
sbottom along with a light stop close-by in mass, both of them being dominantly
L-type. It could also, in general, be noted that since $y_b << y_t$, mixing
between the L- and the R-type sbottom states can at best be modest (when compared
to the stop sector). Thus, the physical sbottoms are nearly pure `chiral' states
and hence their phenomenology is more tractable. Nonetheless, as we will later
see, even such a small mixing could, under certain circumstances, lead to very
interesting phenomenology for the sbottoms.
%
\subsection{The Higgs sector}
\label{subsec:higgs}
%
The Higgs sector of the NMSSM also gets extended nontrivially when compared to 
its MSSM counterpart with the inclusion of a singlet scalar $S$ belonging to
the singlet superfield $\hat{S}$. On electroweak symmetry breaking, CP even
components of the three neutral scalar fields, 
$H_u$, $H_d$ and $S$, mix to give rise
to three CP-even Higgs bosons of the NMSSM. The mass of the SM-like Higgs boson
is given by \cite{Ellwanger:2011sk}
\beq
\label{eq:hmass}
m_h^2 = m_Z^2 \cos^2 2\beta + \lambda^2 v^2 \sin^2 2\beta + \Delta_{\mathrm{mix}}+
\Delta_{\rm rad. corr.} \,
\eeq
with $v = \sqrt{v_u^2 +v_d^2} \simeq 174 \, {\rm GeV}$. 
The first term in the right hand side of \ref{eq:hmass}
gives the tree level squared mass of the SM-like Higgs boson in the MSSM. The
term proportional to $\lambda^2$ is the NMSSM contribution at the
tree level. The term $\Delta_{\mathrm{mix}}$ originates in the so-called
singlet-doublet mixing. In the limit of weak mixing, this is given by
\beq
\label{eq:hmix}
\Delta_{\mathrm{mix}} = \frac{4 \lambda^2 v_S^2 v^2
(\lambda -\kappa \sin 2\beta)^2} {\tilde{m}_h^2-m_{ss}^2}
\eeq
with $\tilde{m}_h^2 = m_h^2 - \Delta_{\mathrm{mix}}$ and 
$m_{ss}^2 = \kappa v_S(A_{\kappa}+4 \kappa v_S)$.
The additional NMSSM contribution at the tree level could enhance the mass of
the SM-like Higgs boson to a level such that to conform to the observed mass,
it does not need to bank much upon the radiative corrections
\cite{Ellwanger:2011sk,Beuria:2015mta}. Hence smaller values of stop masses
(and/or smaller mixing in the stop sector) can be easily afforded thus rendering
the framework more `natural'.

Although the sbottom sector could only play a subdominant role in issues
pertaining to `naturalness', its phenomenology, nonetheless, could significantly
exploit such upshots in the NMSSM stop sector. To conform to the observed mass of
the SM-like Higgs boson, a scenario with light stops in the NMSSM prefers a
genuinely low $\tan\beta \, (\lesssim 5)$
\cite{King:2012is,King:2012tr,Beuria:2015mta}.
As we will discuss in the subsequent sections, this shapes the phenomenology of
the sbottoms in a novel way. It may be reiterated that a light stop which is 
mostly L-type is
accompanied by an L-type sbottom having a close-by mass. An sbottom which is
mostly R-type, however, can be light independent of the masses of the stops.
As we shall see, in both cases, light sbottoms could easily escape experimental
bounds in a simplified NMSSM framework that we adopt in this work.
%
\subsection{The neutralinos and the charginos}
\label{subsec:ewikinos}
The compositions of the neutralinos and the charginos are crucial to the
phenomenology of the sbottoms (and of the stops, as well). While the structure
of the chargino sector of the NMSSM (at the tree level) is identical to that of
the MSSM, the neutralino sector has some essential difference. The difference
arises from the presence of the fermionic component $\widetilde{S}$ corresponding
to the singlet superfield $\hat{S}$ present in the NMSSM superpotential (see
equation \ref{eq:superpot}). The singlino could mix with the higgsinos and the
gauginos.  As we shall see, this could lead to an LSP which has a significant
singlino admixture thus affecting crucially the cascade decay patterns of
the heavier SUSY particles. We assume conserved $R$-parity and hence a stable
LSP.

The symmetric $5 \times 5$ neutralino mass matrix is given by
\beq\label{eq:mneut}
{\cal M}_0 =
\left( \begin{array}{ccccc}
M_1 & 0 & -\dfrac{g_1 v_d}{\sqrt{2}} & \dfrac{g_1 v_u}{\sqrt{2}} & 0 \\
& M_2 & \dfrac{g_2 v_d}{\sqrt{2}} & -\dfrac{g_2 v_u}{\sqrt{2}} & 0 \\
& & 0 & -\mueff & -\l v_u \\
& & & 0 & -\l v_d \\
& & & & 2 \kappa v_S
\end{array} \right) \;
\eeq
in the basis $\psi^0=\{\widetilde{B},~\widetilde{W}^0, ~\widetilde{H}_d^0, 
~\widetilde{H}_u^0, ~\widetilde{S}\}$ \cite{Ellwanger:2009dp,Kraml:2005nx}.
In the above expression, $M_1$ and $M_2$ stand for the soft SUSY-breaking masses
of the $U(1)$ ($\widetilde{B}$) and the $SU(2)$ ($\widetilde{W}$) gauginos,
respectively. The rest of the variables appearing in equation \ref{eq:mneut}
have already been introduced in the text.
The above mass-matrix can be diagonalized by a matrix $N$
\cite{Ellwanger:2009dp,Kraml:2005nx}:
\beq
N^* {\cal M}_0 N^\dagger 
= \mathrm{diag} (\ntrlone, \ntrltwo, \ntrlthree, \ntrlfour, \ntrlfive)
\label{eq:diagN1}
\eeq
such that the five neutralino mass-eigenstates (in order of increasing mass as
`$i$' varies from 1 to 5, in our present study) can be written in a compact
form in terms of the five weak eigenstates ($\psi_j$, with $j=1,....,5$) as
\beq
\chi_i^0 = N_{ij} \psi_j^0
\label{eq:diagN2}
\eeq

As can be gleaned from the entries of the mass matrix, the singlino state mixes
with the gaugino states only via the higgsino sector. Hence these mixings are
rather suppressed. In contrast, the singlino-higgsino mixing is direct and could
be appreciable depending upon the relative values of $\mueff$, $2 \kappa v_S \,
(=m_{\widetilde{S}})$ and the value of $\lambda$. Thus, in a setup where $\mueff$
and $2\kappa v_S$ are small but of comparable magnitudes, the neutralino sector
becomes rather involved with the presence of low-lying neutralino states
(including the LSP) having widely varying singlino admixtures. This can
alter the phenomenology of the light squarks in an essential manner, in particular, 
that of the light sbottoms which is the subject of the present work.

The chargino mass matrix of the NMSSM scenario under consideration, in the basis
\beq
\psi^+ = \left( \begin{array}{c}
                 -i \widetilde{W}^+ \\
                 \widetilde{H}_u^+ 
                \end{array} \right) \quad , \quad
\psi^- = \left( \begin{array}{c}
                 -i \widetilde{W}^- \\
                 \widetilde{H}_d^- 
                \end{array} \right) \quad ,
\eeq
is given by
\beq
{\cal M}_C = \left( \begin{array}{cc}
                    M_2   & \quad  g_2 v_u \\
                 g_2 v_d  & \quad \mueff 
             \end{array} \right) .
\eeq
This differs from ${\cal M}_C$ of the MSSM in the entry $\mueff$ (which is $\mu$
for the MSSM). As in the MSSM, this asymmetric matrix is diagonalized by two
$2 \times 2$ unitary matrices $U$ and $V$:
\beq
U^* {\cal M}_C V^\dagger = \mathrm{diag} (\mcharonepm , \mchartwopm)
\label{eq:uvmatrix}
\eeq
with $\mcharonepm < \mchartwopm$.

Relevant interactions involving the sbottom, the neutralinos, the chargino and
the Higgs boson are discussed in section \ref{subsec:interactions}.
In the present work, we restrict ourselves in a `simplified' setup where the
lighter three neutralinos are typically light and are mixtures of the singlino
and the higgsino states while the lighter chargino is almost purely higgsino.
The heavier states from both the sectors are made to be heavy enough 
by choosing $M_1$ and $M_2$ to be rather large so that
they mostly decouple from the phenomenology.
%
\subsection{Relevant interactions}
\label{subsec:interactions}
In the context of the present work, the important interactions are those
involving the sbottoms, the neutralinos and the charginos. While we would rely
on the strong production modes of the sbottoms (in pairs), the phenomenology
would crucially depend not only on how the sbottoms decay to neutralinos and
the charginos, but also on how these decay products cascade down to the LSP.
Below we outline these interactions in brief.

Interactions of sbottom with neutralinos are of the generic form
\[ b\tilde{b}_i\tilde{\chi}^0_n : \quad 
  g_2 \, \bar{b} \left( a_{in}^{\tilde{b}} P_R + b_{in}^{\tilde{b}} P_L \right)  
  \tilde{\chi}_n^0 \tilde{b}_i + {\mathrm{h.c.}} 
\]
which is similar to the MSSM case except for the fact that the neutralino
index `$n$' now runs from 1 to 5 (see equation \ref{eq:mneut})
\cite{Kraml:2005nx} to include a singlino-like state. $P_{R,L}$ are the standard
projection operators given by ${{1 \pm \gamma_5} \over 2}$. However, couplings
of sbottoms to charginos are identical to those in the MSSM and are given by
\cite{Bartl:2003pd}
\[ t \tilde{b}_i\tilde{\chi}^+_j : \quad 
  g_2 \, \bar{t}\left( l_{ij}^{\tilde{b}} P_R + k_{ij}^{\tilde{b}} P_L \right)  
  \tilde{\chi}_j^+ \tilde{b}_i + {\mathrm{h.c.}} \hskip 10pt  .
\]
In the above expressions, $a_{in}^{\tilde{b}}$, $b_{in}^{\tilde{b}}$,
$l_{ij}^{\tilde{b}}$ and $k_{ij}^{\tilde{b}}$ are all as given in reference
\cite{Bartl:2003pd}. Nonetheless, for future purposes, we write down the
factors $l_{ij}^{\tilde{b}}$ and $k_{ij}^{\tilde{b}}$ (that appear in the
decays of sbottoms to the chargino states) explicitly as follows:
\[ l_{ij}^{\tilde{b}} = 
 -{\cal R}_{i1} U_{j1}+ y_b {\cal R}_{i2} U_{j2},
\qquad
k_{ij}^{\tilde{b}} = y_t {\cal R}_{i1} V_{j2}^* \nonumber
\]
and  $y_b = {m_b \over {\sqrt{2} m_W \cos\beta}}$ and $y_t=
{m_t \over {\sqrt{2} m_W \sin\beta}}$.

The decays of neutralinos and the lighter chargino that involve the Higgs bosons
(in particular, the light, neutral ones) and the gauge ($Z$ and $W$) bosons
are governed by the following interactions.  The Higgs-neutralino-neutralino
coupling is given by
\begin{eqnarray}
H_a \chi^0_i \chi^0_j & : & \frac{\l}{\sqrt{2}} (S_{a1} \Pi_{ij}^{45} +
S_{a2} \Pi_{ij}^{35} + S_{a3} \Pi_{ij}^{34}) - \sqrt{2} \k S_{a3}
N_{i5} N_{j5} \nonumber \\
& & + \frac{g_1}{2} (S_{a1} \Pi_{ij}^{13} - S_{a2}
\Pi_{ij}^{14}) - \frac{g_2}{2} (S_{a1} \Pi_{ij}^{23} - S_{a2}
\Pi_{ij}^{24}) \nonumber
\end{eqnarray}
where $\Pi_{ij}^{ab} = N_{ia}N_{jb}+N_{ib}N_{ja}$ \cite{Ellwanger:2009dp}
and $N$ is given by equations \ref{eq:diagN1} and \ref{eq:diagN2}.
On the other hand, the $Z$-neutralino-neutralino coupling is determined by the
factor \cite{Choi:2004zx}
\begin{eqnarray}
Z \chi^0_i \chi^0_j & : & g_2 \left(N_{i3}N_{j3} - N_{i4} N_{j4} \right)  \nonumber
\end{eqnarray}
while $W^\pm$-chargino-neutralino interaction is determined by the two bilinear
charges ${\cal W}_L$ and ${\cal W}_R$ where \cite{Choi:2004zx}
\[
W^\pm \chi_i^\mp \chi^0_j  :   
\quad 
\left\{ g_2 {\cal W}_{Lij} = g_2 (U^*_{i1} N_{j2} + {1 \over \sqrt{2}} U^*_{i2} N_{j3}),
\quad
{g_2 \cal W}_{Rij} = g_2 (V^*_{i1} N^*_{j2} - {1 \over \sqrt{2}} V^*_{i2} N^*_{j4})
\right\},
\]
$U$ and $V$ being the unitary matrices given by equation \ref{eq:uvmatrix}.
All through, $g_1$ and $g_2$ stand for $U(1)$ and $SU(2)$ gauge couplings, 
respectively.
%
\section{Sbottoms at the LHC Run-I}
\label{sec:sbot-lhc1}
As has been mentioned earlier, given that only a modest mixing between L- and
R-type sbottom states is possible, broad phenomenological studies involving
sbottoms can safely be carried out presuming them to be pure `chiral' states.
Hence in the subsequent discussions, we would systematically take up the cases
of a light sbottom which is either L- or R-type. However, in the light of the
discussion above, we would also demonstrate the impact of even a minuscule
admixture of L-type sbottom in the lighter sbottom state which is otherwise 
dominantly R-type.
%
\subsection{The case with a light $\sbone \equiv \sbleft$}
\label{subsec:sbotl}
%
The published bounds on the mass of $\sbone \equiv \sbleft$ from the LHC Run-I
\cite{Khachatryan:2015wza,Aad:2015pfx} vary depending on how it
decays\footnote{While the present work is being finalized, The ATLAS
Collaboration \cite{Aad:2016tuk} has come up with some tighter lower bounds on
the masses of stop and sbottom squarks from the 13 TeV run of the LHC under
similar sets of theoretical assumptions. This might have some limited numerical
bearing on the present work. We, however, continue to use the analyses of
references \cite{Khachatryan:2015wza,Aad:2015pfx} which are incorporated and
validated in a popular framework like {\tt CheckMATE}
\cite{Drees:2013wra,Kim:2015wza}. The analyses discussed here and the conclusions
drawn thereof are expected to remain broadly unaltered when subjected to the
latest experimental results.}.
As indicated in the Introduction, it touches $\sim 650$ GeV when $\sbleft$ decays
100\% of the times into the $b$-LSP mode. However, in a scenario with relatively
light higgsino-like neutralinos, $\sbleft$ could decay into these states. Unlike
in the MSSM, where these decays would be suppressed since these are driven by the
weak bottom Yukawa coupling (which cannot compete with an analogous setup with,
say, a bino/wino-like LSP neutralino), in the case of the NMSSM these would
dominate. This is precisely since, in our scenario, the LSP is more like a
singlino and thus its coupling to sbottom would be suppressed. Thus, in the
present case, $\sbleft$ could dominantly decay to higgsino-like neutralinos
followed by the latter decaying to the singlino-like LSP. The other possibility,
under such a circumstance, is that $\sbleft$ decays to a higgsino-like chargino
and a top quark. Note that once the phase space for this decay is available,
this could be the dominant two-body decay mode for $\sbleft$ since it is driven
by the top Yukawa coupling. Hence, in any case, the assumption of
BR[$\sbleft \to b$ LSP] would not hold for our scenario and the mass-bound on
$\sbleft$ mentioned above is easily evaded.

It can now be noted that the published bound on the light sbottom mass touches
$\sim 470$ GeV when the two-body decay branching ratios
BR[$\sbone \to t \charonem$] followed by BR[$\charonemp \to W^\mp \ntrlone$]
are both 100\% \cite{Aad:2014pda,Aad:2015pfx}. Such decays can give rise to
same-sign dilepton (SSDL) final states (which are used to put the above bound)
thanks to two sources of $W$-boson in an sbottom cascade; the top quark and the
chargino. An SSDL pair originates from two $W$-bosons of the same sign that
would come from the cascades of the $\tilde{b} \tilde{b}^*$ pair originally
produced in the hard scattering. Being known to be a very clean search channel,
the SSDL final state offers an efficient probe to the sbottom sector. From the
above discussion it is apparent that, in our scenario, this could only be the
case if $\sbone$ is dominantly L-type thus attracting the bound
mentioned above.

Here, we would like to make a very general observation (valid in the MSSM as
well) which is, to the best of our knowledge, new. In a setup with lighter
higgsino-like neutralinos and chargino (the lightest states in case of the MSSM),
there could be a region of phase space where the two-body decay
$\sbone \to t \charonem$ is closed but the three-body decays
$\sbone \to t W^- \ntrlone$ (via off-shell chargino)
and/or $\sbone \to b W^+ \charonem$ (via off-shell top), 
which exploit(s) the same enhanced coupling
$\sbone$-$t$-$\charonem$ could compete and, under favorable circumstances
(at low virtuality), beat the two-body $\sbone \to b \ntrltwothree$ decay rate.
%
\begin{figure}[t]
\centering
\includegraphics[height=0.3\textheight, width=0.49\columnwidth , clip]{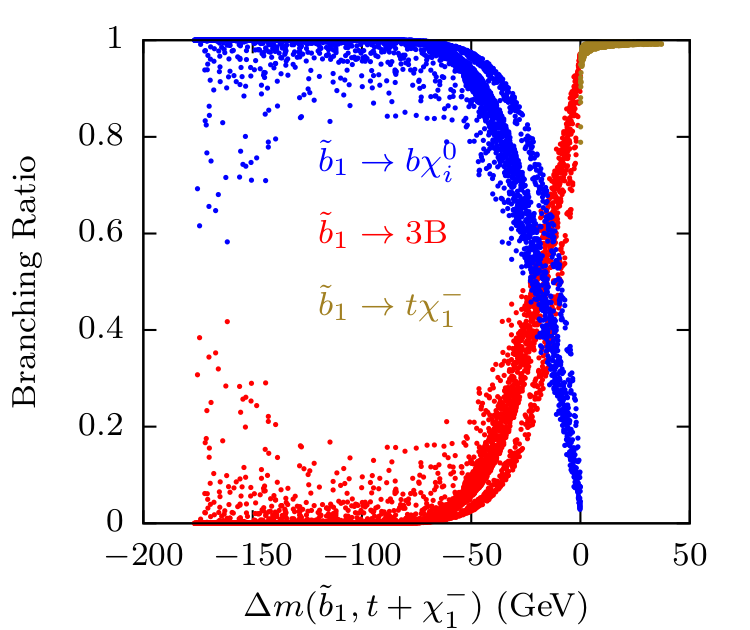}
\includegraphics[height=0.3\textheight, width=0.49\columnwidth , clip]{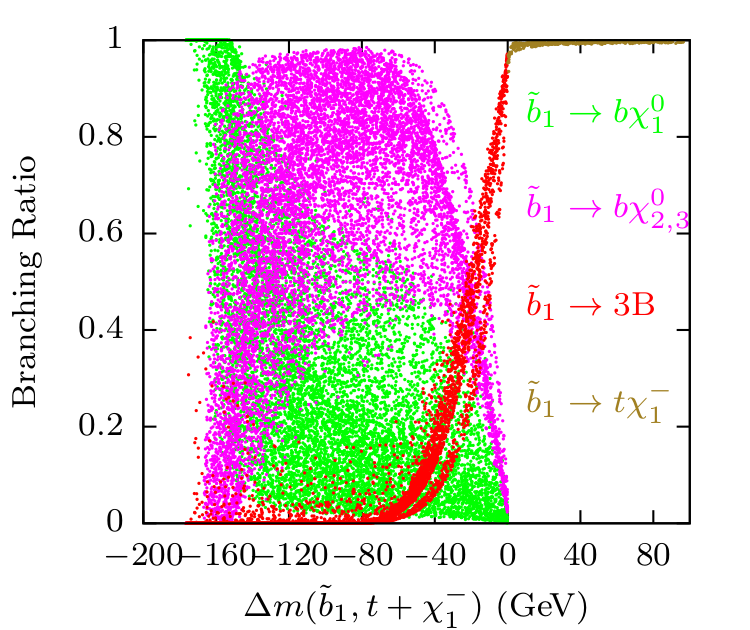}
\vskip 10pt
\includegraphics[height=0.3\textheight, width=0.6\columnwidth , clip]{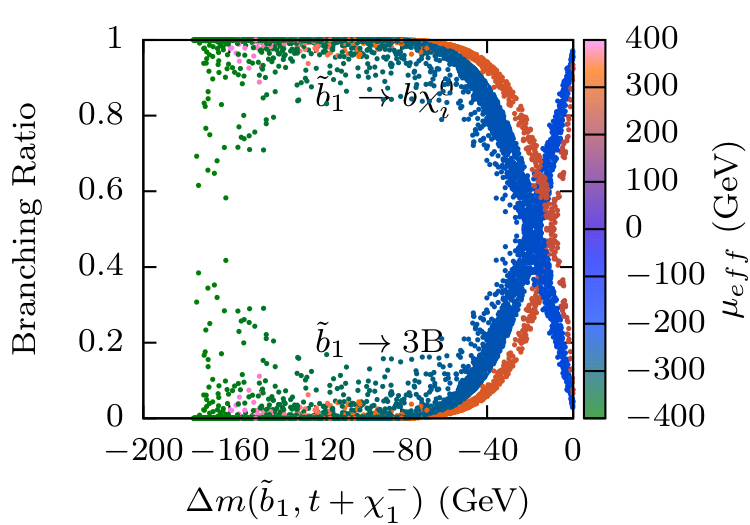}
\caption{
Variations of branching fractions of $\sbone$ into various two- and three-body
decay modes as functions of $\Delta m (\sbone, t+\charonem)$ with $\mueff$ being 
indicated in the color palette. The scan is performed with $1 < \tan\beta < 3$,
$0.6 < \lambda < 0.75$, $0.01 < \kappa < 0.4$, $100 < |\mueff|< 400$ GeV,
$|A_{\lambda,\kappa}|<1$ TeV, $\msQ3=570$ GeV, $\msU3=\msD3=1500$ GeV and
$A_{t,b}=0$, $M_1,M_2=1$ TeV and $M_3=2$ TeV. The value of $m_{\sbone}$ is
fixed around 400 GeV. Only those points are chosen for which the singlino fraction
in the LSP is larger than 50\%. `3B' stands for 3-body decays. See text for further details.}
\label{fig:3b-vs-2b}
\end{figure}

The phenomenon is illustrated in figure \ref{fig:3b-vs-2b} in our NMSSM scenario
with a light sbottom which is dominantly L-type. It shows the variations of decay
branching fractions of $\sbone$ in various 2- and 3-body decay modes as functions
of $\Delta m (\sbone, t+\charonem)$, i.e., the difference of mass of $\sbone$ and
the sum of the masses of the top quark and the lighter chargino.
$\Delta m (\sbone, t+\charonem)$ is varied by varying $\mueff$ which is reflected
in the color palette. The scatter plot on the left of the upper panel shows how
the 3-body decay branching ratio (in red) takes off significantly in advance of
the 2-body threshold of $\sbone \to t \charonem$ (in light brown) and how it
overtakes the collective 2-body branching ratio into $b \ntrlonetwothree$ final
states (in blue). This then indicates the onset of the regime where bounds from
both $\sbone \to b$ LSP and $\sbone \to t \charonem$ cease to exist. The right plot
of the upper panel reveals that the BR[$\sbone \to b \ntrltwothree$] has the
dominant share in the collective 2-body branching ratio over most of the parameter
space. The 2-body branching ratio for the decay $\sbone \to b \ntrlone$ (singlino)
is mostly small and only becomes significant when $\Delta m$ is large negative
thus implying the decays $\sbone \to b \ntrltwothree$ suppressed by phase space.

Our goal is to find how light the sbottom could be under such a circumstance.
We thus choose an sbottom mass of 400 GeV for the illustration. This is
close to but smaller than the  current bound of around 470 GeV obtained assuming
BR[$\sbone \to t \charonem$]=100\%. A good check to ensure that such a relaxation
is indeed possible is to subject the parameter point to a {\tt CheckMATE}
analysis. We find that this is indeed the case. In principle, under such a
circumstance, an even lighter sbottom could have escaped the current experimental
searches.

In any case, it is important to realize that such a light $\sbone \equiv \sbleft$
would imply a nearby light stop. The latter could generically escape the latest
LHC bounds obtained assuming $\stone \to t \chi_1^0$ or $\stone \to b \chi_1^-$
only if it shares its decay branching fractions among available modes
\cite{Aad:2015pfx,Khachatryan:2015wza}. This, in turn, is possible in two ways:
first, if there is some mixing present in the stop sector leading to some admixture
of R-type stop in a predominantly L-type $\stone$; secondly, even for an unmixed
L-type $\stone$ if it could decay to heavier neutralinos.

A careful look at the left plot of the upper panel of figure \ref{fig:3b-vs-2b}
reveals a narrow slit within the red and the blue bands. The plot in the bottom
panel reveals that the slits actually separate two strands which have $\mueff <0$
(in blue) and $\mueff >0$ (in orange). A priori, this may not be unusual given
that the sign on $\mueff$ is known to affect phenomenology in a modest way
\cite{Han:2015tua}.
In fact, we found that the 3-body decay widths of $\sbone$ remain comparable for
either sign of $\mueff$ in the regime under consideration. Rather, the slits have
their origin in the palpably different 2-body decay widths of $\sbone$ for negative
and positive $\mueff$, the latter resulting in a larger width. This effect, in
turn, can be traced back to different couplings of sbottoms to the gaugino components of the
light neutralino states for the two signs on $\mueff$. Note that even though we
work in a regime where $\mueff << M_1= M_2$ (=1.5 TeV), i.e., in a somewhat `deep
higgsino region', a suppressed bottom Yukawa coupling (as is expected for low
values of $\tan\beta$) makes way for even small gaugino admixtures to play the
lead role. Indeed, this effect can be minimized by increasing the values of $M_1$
and $M_2$ further. Also note that, the small values of $\Delta m$ on the left
part of the plot indeed correspond to large $\mueff$ (mostly negative) and hence
to a heavier chargino.

\begin{figure}[t]
\centering
\includegraphics[height=0.15\textheight, width=0.70\columnwidth , clip]{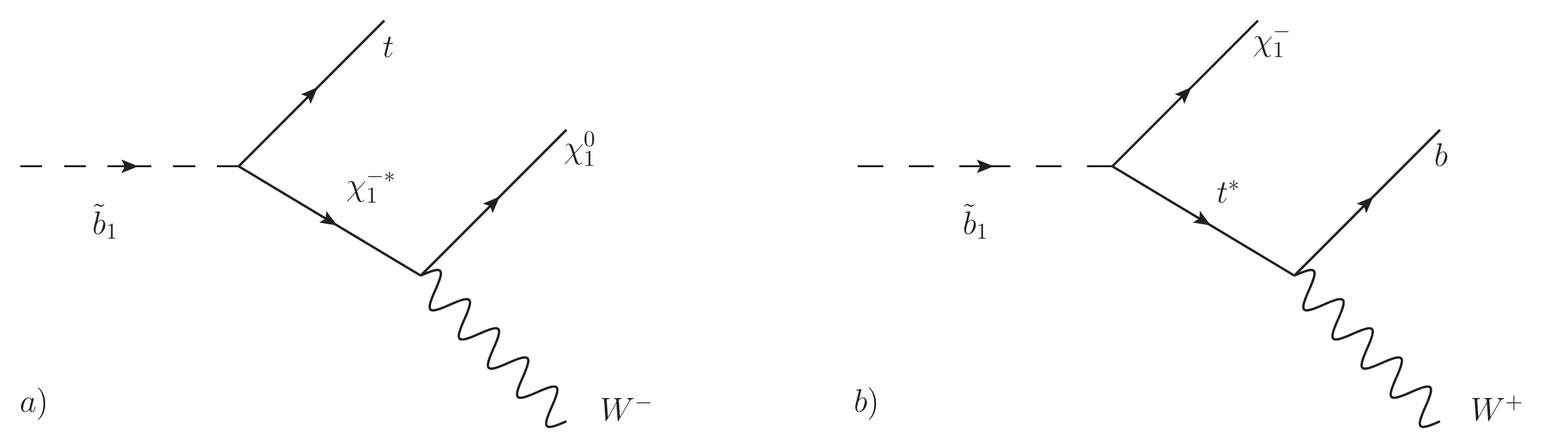}
\caption{Feynman diagrams representing 3-body decays of $\sbone$.}
\label{fig:feyn-diag}
\end{figure}

The Feynman diagrams of the two possible modes of 3-body decay of sbottom are
shown in figure \ref{fig:feyn-diag}: $\sbone \to t W^- \ntrlone$ proceeds
through a virtual $\charonem$ while $\sbone \to b W^+ \charonem$ proceeds
through a virtual top quark. The former is dominant at low values of $\Delta m$
for which the chargino is heavier. The latter dominates for low values of
$\mueff$ where the 3-body final state competes with the 2-body one.
%
%

It turns out, however, that the scattered points in the left part of these plots
with the 3-body decays reaching up to $\sim40$\% cannot 
survive bounds from stop searches at the LHC. This is because for these points 
the LSP is relatively light ($\sim 50$ GeV) and thus BR[$\stone \to t \; {\mathrm{LSP}}$] 
could still be 100\% even though the LSP is mostly singlino-dominated 
(since $\kappa$ is small and $\mueff$ is large) thanks to the large top Yukawa
coupling. Also, since
$\mueff \approx \msbone \approx \mstone \sim 400$ GeV (both $\sbone$ and $\stone$
being dominantly L-type) and thus the chargino is very close to the stop mass,
the 2-body decays of stop to higgsino-like states would be suppressed. Given the 
nature of the couplings there could be a competition among various decay modes of
stop. An appropriate LHC study on the stop sector \cite{Aad:2015pfx} rules these
points out.

In the context of the NMSSM, it is particularly noteworthy that published LHC
results \cite{Aad:2015pfx} do not constrain $\msbone$ at all when
$\mntrltwo \lesssim 280$ GeV for both BR[$\sbone \to b \ntrltwo$] and
BR[$\ntrltwo \to \ntrlone h$] being 100\%. Presence of a singlino-like LSP in
the spectrum and a pair of, rather degenerate, higgsino-like neutralinos could
naturally fulfil such conditions in the cascade decay of $\sbone \equiv \sbleft$
if its decay to the $t\charonem$ final state is not only kinematically disallowed
but $\msbone$ is also significantly below the $t\charonem$ threshold (see section
\ref{subsec:threshold}).
%
\subsection{The case with a light $\sbone \equiv \sbright$}
\label{subsec:sbotr}
\begin{figure}[t]
\centering
\includegraphics[height=0.3\textheight, width=0.49\columnwidth , clip]{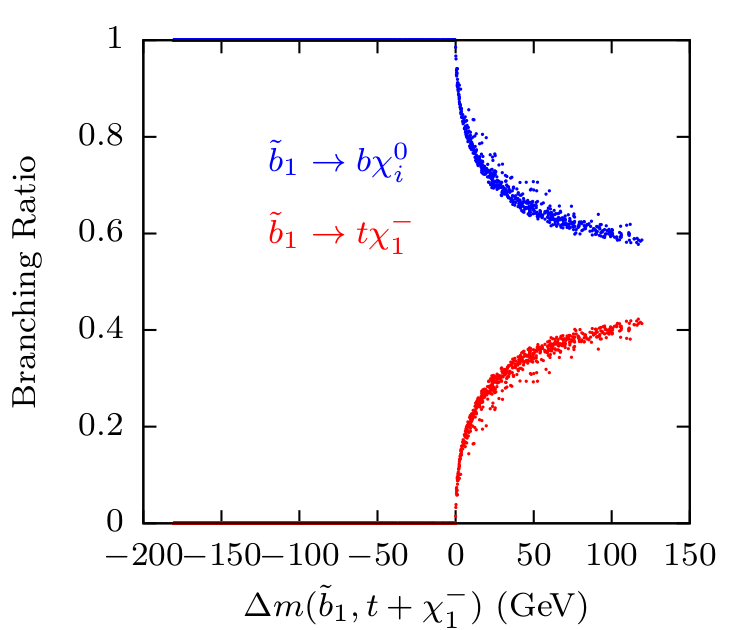}
\includegraphics[height=0.3\textheight, width=0.49\columnwidth , clip]{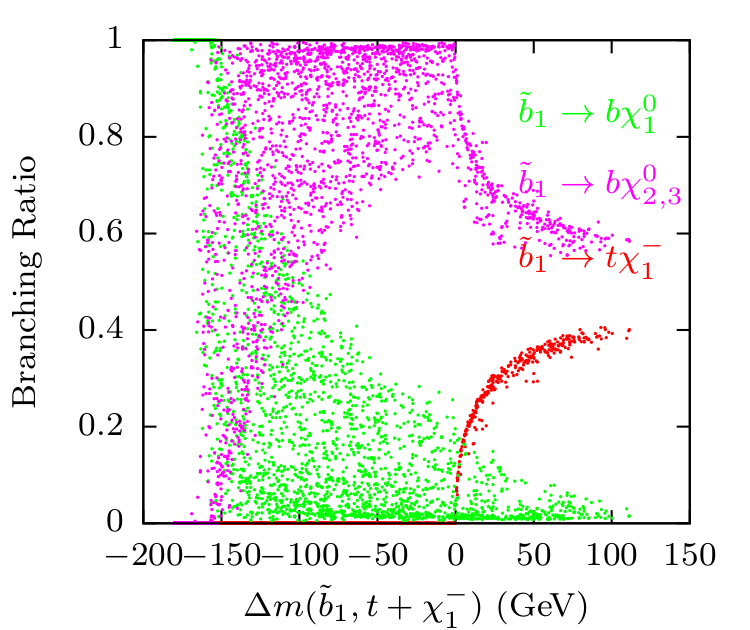}
\caption{Same as in the plots in the upper panel of figure \ref{fig:3b-vs-2b}
but for an R-type $\sbone$.}
\label{fig:sbright-br}
\end{figure}

%
The important difference between the phenomenology of $\sbone \equiv \sbright$
and that of $\sbone \equiv \sbleft$ is that all the relevant decays of $\sbright$
to higgsino-like states are governed by the small bottom Yukawa coupling and thus
they all compete and saturate above the top-chargino threshold. This is shown in
figure \ref{fig:sbright-br}. When compared to figure \ref{fig:3b-vs-2b}, a sharp
contrast is clearly visible in the vicinity of the bottom-chargino threshold and
beyond. Note that while for $\sbleft$ the decay to the chargino final state
quickly dominates and becomes 100\% above the threshold, the same for $\sbright$
saturates to a value of around 40\%.

As a result, in this regime, bound on a light sbottom, which is R-type, obtained
from the top-chargino mode would be much relaxed when compared to the case of an
L-type sbottom. In practice, the amount of such a relaxation can only be estimated
explicitly using a framework like {\tt CheckMATE}. On the theoretical side, under
such a situation, the lower bound on the mass of the $\sbright$ is restricted by
how low one could go down in the chargino mass. The latter is constrained by LHC
analyses involving the chargino-LSP system
when the chargino decays 100\% of the
times to $W$-LSP, which is the case in the NMSSM scenario we are discussing.
References \cite{Aad:2015eda,Khachatryan:2014qwa} indicate that under such a
situation chargino mass as low as 200 GeV is allowed\footnote{{{\footnotesize It has recently been reported in \cite{Xiang:2016ndq} that mass-reaches like
$m_{\chi_{2,3}^0,\, \chi_1^\pm} \sim 320 (500)$ GeV are possible for
the current 13 TeV, 30 fb$^{-1}$ (future 14 TeV, 300 fb$^{-1}$) run of
the LHC. Thus, masses for these excitations that are used in most of the 
benchmark points are within the reach of 13 TeV LHC run.} }} for a LSP mass around 100
GeV.
%
%
\subsection{The case with $\msbone < m_t + \mcharonem$}
\label{subsec:threshold}
%
Issues with sbottom having mass below the top-chargino threshold (large $\mueff$)
are more or less similar for $\sbleft$ and $\sbright$. The important exception is
that for the latter case the 3-body decays are not at all favored below the
top-chargino threshold. In both cases, the two body decays to $b \ntrli$ dominate.
In this regime, it is, however, important to study how the these latter BRs are
shared among $b$-LSP ($\ntrlone$; singlino-like) and $b\ntrltwothree$
(higgsino-like) since these rates determine which kind of experimental constraints
on sbottom mass would be pertinent. It can be noted here that sbottoms are
searched for at the LHC assuming their three pure modes of decays, viz.,
$\sbone \to b \ntrlone$, $\sbone \to b \ntrltwo$ followed by
$\ntrltwo \to h \ntrlone$ and $\sbone \to t \charonem$
\cite{Aad:2014pda,Chatrchyan:2013fea}. For sbottom mass below the top-chargino
threshold the last decay mode simply does not exist.

In the right plot of the upper panel of figure \ref{fig:3b-vs-2b} and in the
right plot of figure \ref{fig:sbright-br} we illustrate these shares of BRs for
$\sbleft$ and $\sbright$, respectively. As before, the outer edges of the green
and the pink regions correspond to the maximum admixture (50\%) of higgsino in
the LSP that we allow for in this analysis. The figures show that BR in bottom-LSP
drops sharply as $\mueff$ decreases (as we go from left to right) since this
opens up the decays to $b \ntrltwothree$. Clearly, thus, the bounds on sbottom
mass that assume BR[$\sbone \to b \,$ LSP]=100\% would not hold in this regime.
The next regime with BR[$\sbone \to b \ntrltwothree$]=100\%, where sbottom
searches put bound on its mass, also gets affected because of such a sharing of
branching fractions thus resulting in a relaxed bound. The situation is a bit
worse for $\sbleft$ for which the 3-body decays also get their shares in this
regime.
%
\subsection{Effect of mixing in the sbottom sector}
\label{subsec:mixing-effect}
The discussions in section \ref{subsec:sbotl} could well point to how things
change drastically even for a small admixture of $\sbleft$ in an otherwise
$\sbright$-dominated $\sbone$. It may be reiterated that, unlike in the case of
the MSSM, one could afford a significantly low $\tan\beta$ ($\sim 2$) in the
NMSSM when $\lambda$ is large. As a consequence, $y_t \over y_b$ could become
large. Given that the mixing in the sbottom sector is naturally restrained, it
is all the more interesting to observe how significantly this affects the decay
pattern of $\sbone$ having only very small $\sbleft$ admixture.

\begin{figure}[t]
\centering
\includegraphics[height=0.25\textheight, width=0.45\columnwidth , clip]{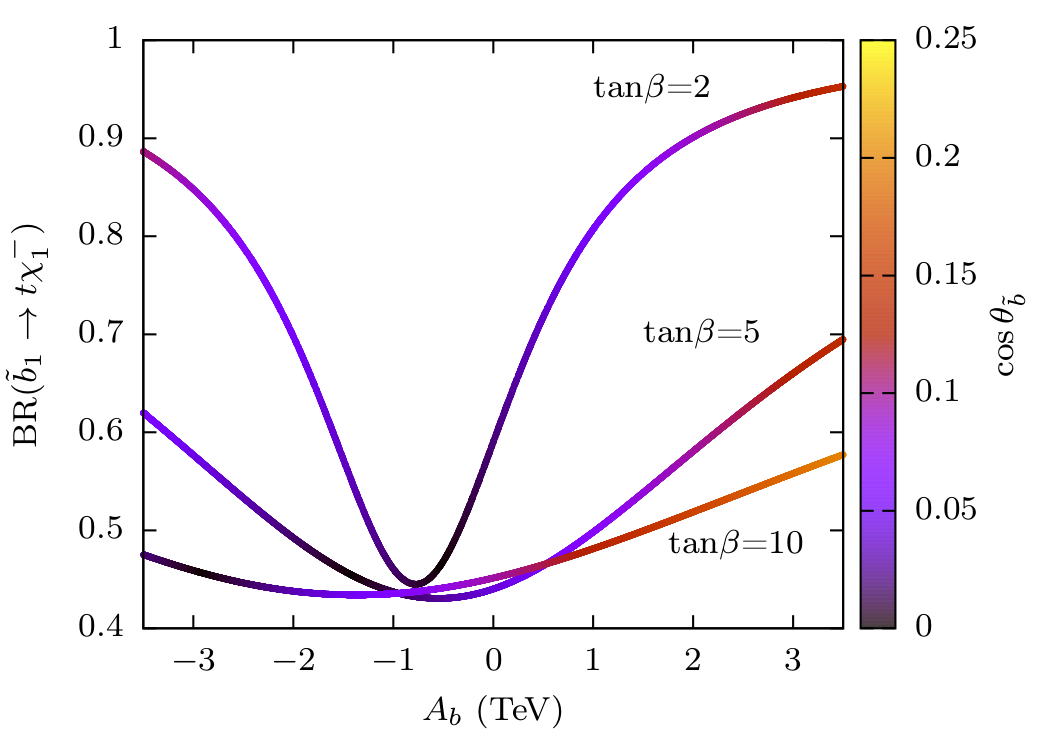}
\includegraphics[height=0.25\textheight, width=0.45\columnwidth , clip]{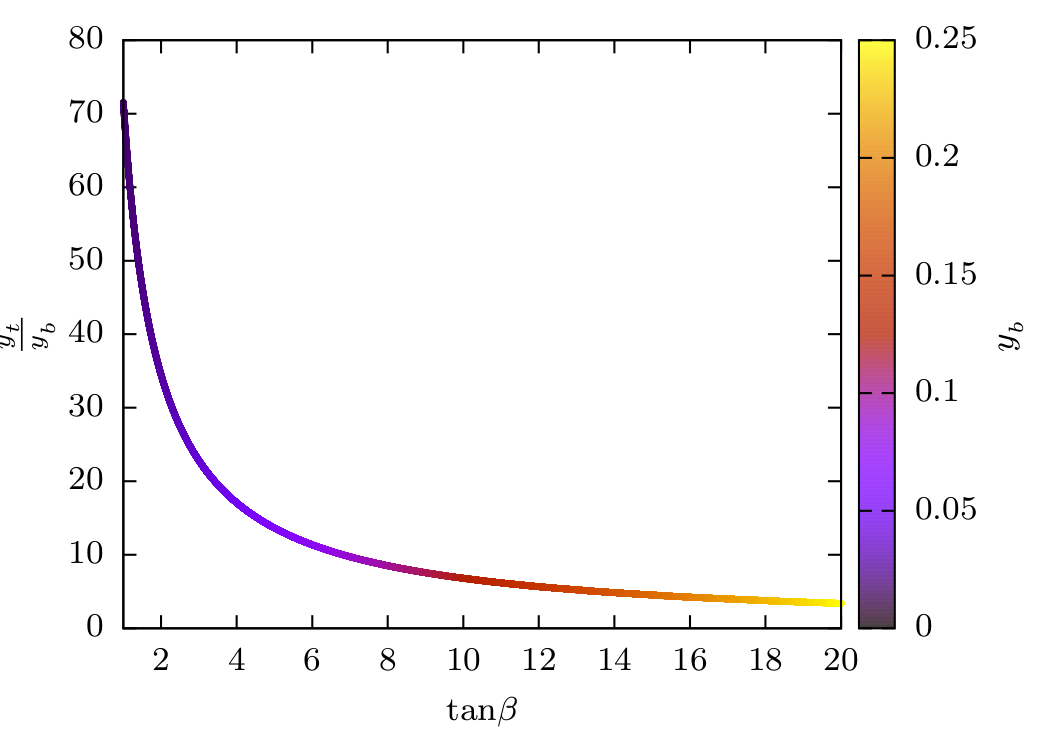}
\caption{
{\bf Left:} Variations of BR($\sbone \to t \charonem$) as a function of $A_b$ for
various values of $\tan\beta$. The color palette indicates the varying magnitude
of sbottom mixing along each curve in terms of $\cos\theta_{\tilde{b}}$. Other
parameters are fixed as follows:
$\mueff=-300$ GeV, $m_{\sbone} \approx 700$ GeV, $m_{\sbtwo} \approx 750$ GeV,
$\msQ3=680$ GeV, $\msU3=1.5$ TeV, $\msD3=630$ GeV, $A_t=0$, $M_1=1$ TeV,
$M_2=1.5$ TeV and $M_3=2$ TeV.
{\bf Right:} Variation of the ratio $y_t \over y_b$ as a function of $\tan\beta$.
The color palette indicates the varying magnitude of $y_b$ along the curve.
}
\label{fig:br-ab-cosb}
\end{figure}

In the left plot of figure \ref{fig:br-ab-cosb} we illustrate the implication of
such a small $\sbleft$-admixture in $\sbone$ for the decay branching fraction
BR[$\sbone \to t \charonem$]. We take close values of $\msQ3$ ($\sim 680$ GeV)
and $\msD3$ ($\sim 630$ GeV) which facilitate mixing in the sbottom sector.
Also, we fix $\mueff$ at $-300$ GeV, a value which is representative of a
`natural' setup.  Variations of colour along the individual curves reflect the
varying chiral-mixing (in terms of $\costhetab$) which can be gleaned from the
colour palette. Note that $\costhetab =0$ implies a pure R-type $\sbone$ (see
section \ref{subsec:sbot-stop}).

Variations of BR[$\sbone \to t \charonem$] are then shown for three representative 
values of $\tan\beta$ ranging over small to intermediate values. It can be seen 
from the figure that for all values of $\tan\beta$, the said branching fraction 
increases with increasing $|A_b|$. This is expected since increasing $A_b$ can 
indeed induce larger $\sbleft$ admixture in $\sbone$ which is predominantly 
R-type in our present setup. The positions of the troughs in all these curves are 
primarily determined by the values of $A_b$ for which the $\sbleft$ admixtures 
are the smallest.
For small $\tan\beta \, (\approx 2)$, this can be clearly seen in the figure. The
relative locations of the troughs for curves with different values of $\tan\beta$,
however, have complicated dependence on both $A_b$ and $\tan\beta$. Also, one
could find that, for large values of $|A_b|$, BR[$\sbone \to t \charonem$] could
get as large as 95\%, 70\% and 55\% for $\tan\beta=2,5$ and 10, respectively with
a small to moderate mixing.
Furthermore, one can see that the gradients of the variations are steeper for
smaller $\tan\beta$ values. This can be seen as a result of how the ratio
$y_t \over y_b$ varies as a function of $\tan\beta$. This variation is shown
in the right plot of figure \ref{fig:br-ab-cosb}.

Such large branchings of $\sbone$ to top quark and chargino, under the present
setup, are definite outcomes of possible low values of $\tan\beta$ which can be
easily accommodated in a `natural' NMSSM scenario. In contrast, in the MSSM, such
a large branching fraction for $\sbone \to t \charonem$ is rather difficult to
attain since the observed mass of the SM-like Higgs boson restricts $\tan\beta$
from the bottom.
%
%
\section{Benchmark scenarios and the 13 TeV LHC}
\label{section:benchmark-lhc13}
As has been already pointed out, studying the phenomenology of sbottoms is
important in its own right and could shed crucial light on the physics of the
squarks from the third generation by complementing/supplementing the studies
in the stop sector. The subject gets further involved in nontrivial ways in a
scenario like `natural' NMSSM, in particular, in the light of available results
from the LHC Run-I. A systematic first study into this should be based on some
salient aspects that the sbottom sector possesses in such a setup.  Some such
issues are discussed in the previous section.
In this section we first present a few benchmark points which are representative
of the situations described in sections \ref{subsec:sbotl}, \ref{subsec:sbotr} and
\ref{subsec:mixing-effect}. We then study the collider aspects of these scenarios
at the ongoing LHC-13.
%
\subsection{Benchmark scenarios}
\label{subsec:benchmarks}
In table \ref{tab:benchmarks} we present six benchmark scenarios/points (BPs).
For all of them, we fix the gaugino soft mass parameters to be $M_1=$ 1 TeV,
$M_2$= 1.5 TeV and $M_3$ =  2 TeV such that the heavier neutralinos and the
heavier chargino are absent in the cascades of a much lighter $\sbone$
(a simplified scenario with only singlino- and higgsino-like lighter neutralinos
and a higgsino-like lighter chargino). Large values of $\lambda$ ($\gtrsim 0.6$)
are chosen in order to enhance the tree level NMSSM contribution to the 
SM-like Higgs boson mass. As discussed earlier, this requires low values of 
$\kappa$ to ensure the absence of Landau pole below the GUT scale which, in
turn, results in an LSP which is singlino-like. We thus choose 
$\kappa\lesssim 0.25$. Except for BP6, all other benchmark points have 
$A_t=0$, as discussed in section \ref{subsec:sbotl}.

Two common but salient features of these scenarios are that they all have small
$\tan\beta \, (\sim 2)$ and that the LSP is mostly a singlino ($\sim 75\% - 90\%$),
albeit with a non-negligible higgsino admixture. Such a composition is achieved by
choosing relatively large values of $\lambda (\gtrsim 0.6)$ while keeping
$\mueff$ somewhat smaller. Note that such a setup generically guarantees a low
fine-tuning (i.e., more `natural') and singlino-domination in the LSP can thus
be seen to be directly connected to this fact. Palpable mixing among the higgsino
and the singlino states is also unavoidable for large $\lambda$ and since
$\mueff \approx \lambda v_S$ is a possibility.
The benchmark scenarios presented here are checked for their viability via
{\tt CheckMATE} against the available analyses of the LHC Run-I data. These
points also satisfy all Higgs-related constraints and other phenomenological
bounds that are built-in in {\tt NMSSMTools-v4.8.2}
\cite{Ellwanger:2004xm, Ellwanger:2005dv, Ellwanger:2006rn, Das:2011dg,
Muhlleitner:2003vg}.
However, constraints pertaining to muon $g-2$ are ignored while in the dark
matter sector only the upper bound on the relic density ($\Omega_c h^2 < 0.131$)
as incorporated in {\tt NMSSMTools} is respected.

We have presented three benchmark scenarios (BP1-BP3) having $\tilde{b}_1$ which
is purely $\sbright$ while for the scenarios BP5 and BP6 it is purely $\sbleft$.
In BP4 we allow for a little admixture of $\sbleft$ in $\sbone$. Scenarios BP1
and BP2 differ in the masses of $\sbone$ and in the fact whether $h_1$ or $h_2$
is the SM-like Higgs boson. BP2 and BP3 are rather similar except for the
possibility of an additional decay of $\ntrltwothree$ to $h_1$ that is available
for the latter. For these three benchmark points we set 
$m_{\tilde{Q}_3}=m_{\tilde{U}_3}=1500$ GeV\footnote{{\footnotesize Choice of stops and an sbottom as heavy as 1.5 TeV, whenever 
      possible, is made just to ensure that they effectively decouple for our 
      purpose while the other sbottom may still remain light in the scenario 
      under consideration. 
      It may, however, be noted that stops with mass $\lesssim 1$ TeV would 
      not affect yields in the final states we consider and, hence our results, 
      in any significant way.
      As discussed in the Indroduction, relatively heavier stops may not 
      necessarily be in conflict with `naturalness' when $|\mu_{eff}|$ is small 
      enough. This has been checked explicitly by using NMSSTools where the 
      related finetuning parameter yields values $\sim 10$, which is indicative 
      of a healthy naturalness, for all the benchmark points with heavy stop(s).}} and appropriately small values of
$m_{\tilde{D}_3}$ are chosen.

In BP4 we present a scenario where a small admixture of $\tilde{b}_L$ in
$\sbone \, (\simeq \sbright)$ which is achieved by making $A_b$ large even as
$m_{\tilde{Q}_3} \sim m_{\tilde{D}_3}$. As pointed out earlier, even a tiny
left admixture ($\lesssim 2\%$), when assisted by large $\frac{y_t}{y_b}$ (driven
by small $\tan\beta$), could result in a very large branching fraction for
$\sbone \to t \charonep$ (reaching $\sim 90\%$). This is something not usually expected
of $\sbone \equiv \sbright$, in particular, when other two-body decay modes
(to light neutralinos) are kinematically accessible to it.

In BP5 $\sbone$ is made to resemble $\sbleft$ by lowering the $m_{\tilde{Q}_3}$
(=500 GeV $<<\msU3=\msD3=1500$ GeV). BP6 is a point representative of the
cross-over region for the curves that illustrate the variations of the 2-body and
3-body branching fractions of $\sbone$ in figure \ref{fig:sbright-br}. Note that,
in this case, the lighter stop has a mass of around 400 GeV. As pointed out in
section \ref{subsec:sbotl}, this could escape the latest bound from the LHC only
if the stop sector is attributed with some mixing.
Also, BP6 features sub-TeV $\stone$ and $\sttwo$ whose origins are discussed in
section \ref{subsec:sbotl}.

As far as the cascades of the light neutralinos and the lighter chargino are
concerned, it can be seen from table \ref{tab:benchmarks} that the former decay
predominantly to $\ntrlone Z$ followed by $\ntrlone h$ (SM-like) while the latter
decays 100\% of the time to $\ntrlone W^\pm$. Scenario BP5 is an exception for
which BR[$\ntrltwo \to \ntrlone h$] dominates over BR[$\ntrltwo \to \ntrlone Z$].
For all the benchmark points, current bounds on the chargino-neutralino and the
third generation squark sectors are respected, using the public software
{\tt CheckMATE} wherever dimmed necessary.
%

\begin{table}[t]
\begin{center}
  \begin{tabular}{|c||c|c|c|c|c|c|}
   \hline
 \cline{2-7}
 Parameters  & BP1 & BP2 & BP3 & BP4 & BP5  & BP6\\
   \hline
 $\lambda$             & 0.63 & 0.65 & 0.65 & 0.67  & 0.68 & 0.70\\
 $\kappa$              & 0.15 & 0.20 & 0.17 & 0.20  & 0.19 & 0.25\\
 $A_{\lambda}$  (GeV)  & 435 & -600 & -600 & -580  & 660 & -320\\
 $A_{\kappa}$   (GeV)  & -50 & 50 & 50 & 50 & -50 & 85\\
 $\mu_{_{eff}}$ (GeV)  &  200 & -300 & -300 & -300 & 350 & -250\\
 $\tan\beta$           &    2 &    2 &    2 & 2 & 2 & 1.6\\
 $A_t$          (GeV)  & 0 &  0 & 0   & 0 & 0 & 300\\
 $A_b$          (GeV)  & -1000 & -1000 & -1000 & -2500 & 0 & 0\\
 $m_{\tilde{Q}_{3}}$ (GeV) & 1500 & 1500 & 1500 & 550 & 500 & 260\\
 $m_{\tilde{U}_{3}}$ (GeV) & 1500 & 1500 & 1500 & 1500 & 1500 & 400\\
 $m_{\tilde{D}_{3}}$ (GeV) & 330 & 440  & 440 & 520 & 1500 & 1500\\
   \hline
   \hline
 Observables & BP1 & BP2 & BP3 & BP4 & BP5 & BP6\\
   \hline
  Singlino & 0.78 & 0.87 & 0.90 & 0.87 & 0.89 & 0.77\\
   fraction in the LSP    & & & & & &\\
\hline
  $\tilde{b}_R$  fraction in $\sbone$ & 1 & 1 & 1 & 0.98 & 0 & 0\\
   \hline
  $m_{h_1}$ (GeV) & 107.1  &  124.4  &  125.7 & 124.6 & 123.2 & 125.9\\
  $m_{h_2}$ (GeV) & 125.2  &  190.2  &  163.8 & 184.4 & 202.6 & 173.1\\
 $\msbone$ (GeV) & 407.7  & 506.0 & 506.0 & 598.3 & 588.5 & 405.0\\
 $\msbtwo$ (GeV) & 1553.3 & 1552.4 & 1552.4 & 633.9 & 1557.6 & 1549.0\\
 $\mstone$ (GeV) & 1552.8 & 1555.1 & 1555.1 & 650.0 & 605.6 & 396.2\\
 $\msttwo$ (GeV) & 1567.4 & 1564.5 & 1564.5 & 1556.7 & 1557.2 & 533.6\\
 $m_{_{\chi_1^0}}$ (GeV) & 113.7 & 200.1 & 171.7 & 194.3 & 209.5 & 195.1\\
 $m_{_{\chi_2^0}}$ (GeV) & 212.0 & 322.0 & 319.6 & 319.5 & 356.0 & 264.0\\
 $m_{_{\chi_3^0}}$ (GeV) & 238.3 & 328.2 & 329.2 & 327.1 & 376.0 & 274.4\\
 $m_{_{\chi_1^-}}$ (GeV) & 201.0 & 309.2 & 309.2 & 306.4 & 348.8 & 250.0\\
   \hline
 BR($\sbone \to t \chi_1^-$)        & 0.29 & 0.30 & 0.30 & 0.93 & 1 & 0\\
 BR($\sbone \to b \chi_1^0$)        & 0.02 & 0.01 & 0 & 0 & 0 & 0.18\\
 BR($\sbone \to $3-body)            &  0  &0 & 0 & 0 &0& 0.57 \\
 BR($\sbtwo \to t \chi_1^-$)        & 0.73 & 0.76 & 0.76 & 1 & 0 & 0 \\ 
\hline
 BR($\chi_2^0 \to \chi_1^0 Z$)       & 1 & 1 & 0.65 & 0.85 & 0.40 & 0.95\\
 BR($\chi_2^0 \to \chi_1^0 h$)       & 0 & 0 & 0.35 & 0.15 & 0.60 & 0\\
 BR($\chi_3^0 \to \chi_1^0 Z$)       & 1 & 0.99 & 0.79 & 0.96 & 0.80 & 0.98\\
 BR($\chi_3^0 \to \chi_1^0 h$)       & 0 & 0.01 & 0.21 & 0.04 & 0.20& 0\\
   \hline
  \end{tabular}
\end{center}
\caption{Benchmark points studied in the present work.
The slepton and the first two generation squark masses are all
set to 1.5 TeV and the $SU(3)$ gaugino mass, $M_3$ is set to 2
TeV. The $SU(2)$ gaugino mass $M_2$ is set to 1.5 TeV and $U(1)_Y$ gaugino mass
$M_1$ is set to 1 TeV. Remaining sbottom branching fraction in each case is 
attributed to its decays to $b\ntrltwothree$. See text for details.}
\label{tab:benchmarks}
\end{table}
\subsection{Sbottoms at the LHC-13}
%
\begin{table}[h]
\begin{center}
  \begin{tabular}{ |c|c|c|c| }
    \hline
Cascade modes& Intermediate products& Partonic final states \\\hline
 $\sbone \rightarrow b \ntrltwothree, \; \; \sbone \rightarrow b 
 \ntrltwothree$ &  $2b+2(Z/h)$ &  $4\ell+2b+\slashed{E}_T$ \\ 
\hline
 $\sbone \rightarrow t \chi_1^-, \;\; \sbone \rightarrow b \ntrltwothree$
 &  $2b+W^+W^-+(Z/h)$ 
 & $(3\ell, 4\ell)+2b+\slashed{E}_T$  \\ 
\hline
 $\tilde{b}_{1,2} \rightarrow t \chi_1^-, \; \; \tilde{b}_{1,2} \rightarrow t \chi_1^-$ & 
 $2b+2(W^+W^-)$   
 & (SSDL, $3\ell, 4\ell$) $+2b+\slashed{E}_T$  \\ 
\hline
 $\tilde{t}_{1,2} \rightarrow t \ntrltwothree,
 \; \; \tilde{t}_{1,2} \rightarrow t \ntrltwothree$ &  $2b+W^+W^-+2(Z/h)$ 
 & $(3\ell, 4\ell)+4b+\slashed{E}_T$  \\
\hline
 $\tilde{t}_{1,2} \rightarrow t \ntrltwothree,
 \; \; \tilde{t}_{1,2} \rightarrow b \charonep / t \ntrlone$ &  $2b+W^+W^- + Z/h$ 
 & $(3\ell, 4\ell)+2b+\slashed{E}_T$  \\ \hline
\end{tabular}
\end{center} 
\caption{Possible decay modes of $\sbonetwo$, $\stonetwo$. The decays
$\ntrltwothree \rightarrow \chi_1^0 Z/h$ and $\chi_1^- \rightarrow W^- \chi_1^0$
have 100\% branching fraction.}
\label{tab:final-topology}
\end{table}
\begin{table}[ht!]
\begin{center}
\vspace{0.8cm}
\begin{tabular}{ |c|c|c| }
\hline
 Channel & Search channel & Dominant \\  
      ID &                & backgrounds \\
\hline   
  & & \\
 $\mathrm{SRSSDL}$  & $\mathrm{SSDL} + \geq 2j (2b) + \slashed{E}_T$
 & $ t\bar{t}$, $t\bar{t}Z$, $t\bar{t}W$ \\
  & & \\
 $\mathrm{SR3\ell 2b}$  & $3\ell + \geq 4j(2b) + \slashed{E}_T$ 
 & $ t\bar{t}$, $t\bar{t}Z$, $t\bar{t}W$    \\
  & & \\             
 $\mathrm{SR4\ell 1b}$  & $3\ell + \geq 2j(1b) + \slashed{E}_T$ 
 & $t\bar{t}Z$, $t\bar{t}W$   \\
  & & \\
    \hline
  \end{tabular}
\end{center}
\caption{
Classification of signal regions in terms of the actual search channels undertaken
in the present analysis. Leptons have their origins in the $Z$- and the $W$-bosons.
The last column presents the dominant SM background processes corresponding to each
final state. These are inclusive of two hard jets except for the $ZW$ and
$t \bar{t}$ processes for which three-jet inclusive samples are used.
}
\label{tab:signals}
\end{table}
 
In table \ref{tab:final-topology} we present the possible decay modes 
of the sbottoms and the stops and indicate the intermediate products in the 
cascades and the partonic final states thereof.
It may be apparent from the table that a simultaneous analysis involving 
various multi-lepton ($n_\ell \geq $3) and SSDL final states with low $b$-$jet$ 
multiplicity may be a useful probe to such a scenario.
The cascade decays of stops could yield a larger $b$-$jet$ multiplicity. 
This might eventually help recognizing the
presence of stops. In all these cases the leptons have their origins in the
decays of the SM gauge bosons. The cascade of higgsino-like neutralino will vary
crucially determine the signal topology. Identifying $b$-$jets$ will certainly
be helpful in this specific scenario.  Table \ref{tab:cuts} describes suitable
cuts for the above mentioned signal topologies.

The SSDL signal arises only from the cascade $\sbone \to t \chi_1^-$ followed by
$\chi_1^- \to \chi_1^0 W^-$. The pure L-type stop quark will never give an SSDL
signal in this set up. So the observation of the SSDL signal would definitely point
to a $\sbone$ cascade. But as we can see from BP4, due to the effect of large
$\frac{y_t}{y_b}$, it may not be easy to estimate the mixing in the
sbottom sector.
%
\subsection{Simulation setup and selection of final states}
The lowest order (LO) parton-level signal events are generated using
{\tt MadGraph\_aMc@NLO v2.3.3} \cite{Alwall:2014hca}. Background events 
are obtained using {\tt MadGraph\_aMc@NLO v2.1.2} \cite{Alwall:2014hca}. 
In both cases, parton distribution 
function {\tt nn23lo1} \cite{Ball:2014uwa}, default to {\tt MadGraph}, 
is used with the
factorization/renormalization scale set at the default MadGraph setting of
transverse mass. For the signal, the next-to-leading-order (NLO) cross sections
are estimated via {\tt Prospino v2.1} \cite{Beenakker:1996ed}. For the backgrounds,
to take into the account higher order effects, we employ a flat $K$-factor of 1.6
for the inclusive $t\bar{t}$ samples and 1.3 for the rest.

Signal events are showered with {\tt Pythia v8.2} \cite{Sjostrand:2014zea}.
Background events are showered with {\tt Pythia v6.426} \cite{Sjostrand:2006za}
embedded in the MadGraph setup. For background events, we employ the MLM
\cite{Mangano:2006rw, Alwall:2007fs} scheme for matching jets in order to avoid
double counting in the presence of hard partonic jets and parton shower.
Background events are produced with up to three inclusive jets.

Subsequently, both signal and background events are subjected to detector
simulation via {\tt DELPHES v3.2.0} \cite{Selvaggi:2014mya} which includes
{\tt Fastjet v3.1.0} \cite{Cacciari:2011ma} for jet reconstruction. For merging
of jets, we employ the anti-$k_T$ algorithm with cone size set to 0.4 and
require a minimum $p_T^{jet}$ of 20 GeV with pseudorapidity in the range
$|\eta_{jet}| < 2.5$. $b$-tagging efficiency is set to 70\%. Furthermore, we
consider the probability of a $c$-$jet$ and a light quark jet being tagged as a
$b$-$jet$ to be 20\% and 1\%, respectively.

Leptons (electrons and muons) are reconstructed with minimum $p_T^\ell$ of 10
GeV and with $|\eta_\ell| < 2.5$. Leptons having neighbouring (reconstructed)
jets lying within a cone of $\Delta R \le 0.2$ about them are rejected. To
increase the purity of electrons further, we require the ratio of total $p_T$ of
the stray tracks within the cones of their identification to their own $p_T$'s
is less than 0.1. For muons the maximum total $p_T$ of other tracks is required
to be below 1.8 GeV.
\begin{table}[!htbp]
 \begin{center}
  \begin{tabular}{ ||c|c|c|c|| }
    \hline \hline
    Variables & $\mathrm{SRSSDL}$ &  $\mathrm{SR3\ell2b}$ & $\mathrm{SR4\ell1b}$  \\ 
\hline \hline
    $n_\ell$ & 2 (SSDL) &  3 & 4 \\ \hline
    $n_{jet}$ & $\geq 2$ &  $>=4$ & $>=2$  \\ \hline
   $n_{b\text{-}jet}$ & $>=2$  &  $>=2$ & $>=1$  \\ \hline
 $p_T^{j_{(n)}}$ (GeV)  & 
\multicolumn{3}{c||}{$p_T^{j_{(1,2,3,\geq 4)}}> (30, 30, 30, 20)$} \\
 $p_T^{b\text{-}jet(n)}$ (GeV)  & 
\multicolumn{3}{r||}{$p_T^{b\text{-}jet(1,2)} > (40,30)$}\\ \hline
 $p_T^\ell$ (GeV)  & \multicolumn{3}{c||}{$p_T^{\ell (1,2,3,4)} > (30,20,15,15)$} \\ \hline
 $\slashed{E}_T$ (GeV)  & \multicolumn{3}{c||}{$> 100$} \\ \hline
 $\mttwo$ (GeV) & $>$ 90  & -& - \\ \hline
 $H_T$ (GeV)  & $>$ 400 & $>$ 500  & $>$ 500   \\ \hline \hline
\end{tabular}
\caption{
Definitions of the signal regions (SR) indicating the final states they represent
and the sets of selection cuts on the physics objects that are independent of the
benchmark scenarios. By leptons only electrons and muons are referred to. Other
notations follow standard conventions.
}
\label{tab:cuts} 
\end{center}
\end{table}   

In table \ref{tab:cuts} we define three distinct signal regions namely
$\mathrm{SRSSDL}$, $\mathrm{SR3\ell 2b}$ and $\mathrm{SR4\ell1b}$ and the signal
selection cuts used for them in the framework of {\tt MadAnalysis 5v1.1.12}
\cite{Conte:2012fm,Conte:2014zja,Dumont:2014tja} . Jets and leptons are
$p_T$-ordered with the hardest jet being denoted as $j_1$and the hardest lepton
as $\ell_1$. To have a better handle on some important backgrounds like 
$t\bar{t}+jets$, we employ standard kinematic variables like
$H_T= \sum_{_{jets}} |p_T^j|$ and $\mttwo$ \cite{Cheng:2008hk}, where
\bea
\mttwo(p_T^{\ell_1},p_T^{\ell_2},\slashed{p}_T)
=\min_{ \slashed{p}_{T,1}+\slashed{p}_{T,2}
= \slashed{p}_T}\{\max\{m_T( p_T^{\ell_1}, \slashed{p}_{T,1}),m_T( p_T^{\ell_2}, 
\slashed{p}_{T,2})\}\}
\eea
with
\bea
m_T(p_T^\ell,\slashed{p}_T)=\sqrt{2p_T^\ell \slashed{p}_T(1-\cos\phi_{\ell,\slashed{p}_T})}
\eea 
\begin{figure}[t]
\centering
\includegraphics[height=0.22\textheight, width=0.49\columnwidth , clip]{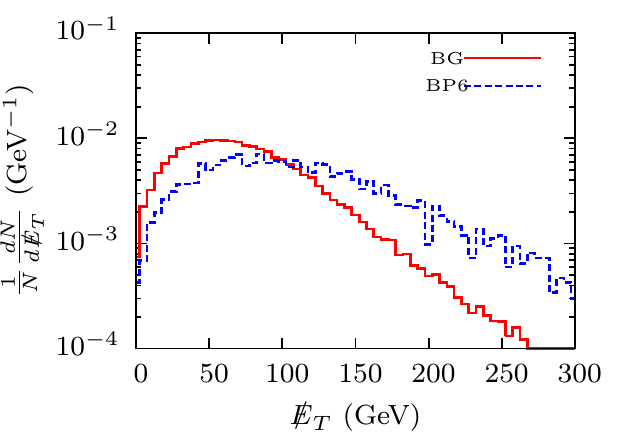}
\includegraphics[height=0.22\textheight, width=0.49\columnwidth , clip]{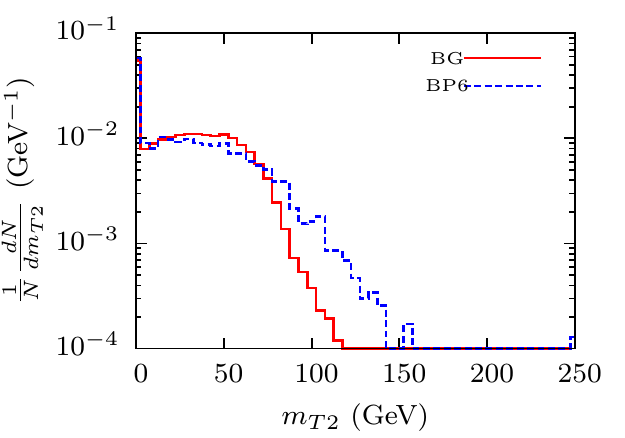}
\vskip 10pt
\includegraphics[height=0.22\textheight, width=0.49\columnwidth , clip]{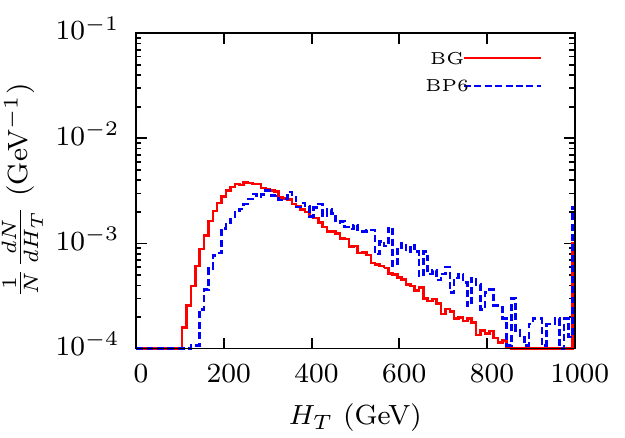}
\caption{
Differential distributions for the kinematic variables $\etmiss$, $\mttwo$ and 
$H_T$ for both signal (in scenario BP6) and the background at LHC-13 for
inclusive final states. See text for details.
}
\label{fig:kinematic-distribution}
\end{figure}
%
\subsection{Results and discussion}
In this subsection we discuss the results of our simulations for the LHC-13
in different signal regions described in table \ref{tab:cuts}. These signal
regions satisfy a general set of cuts on lepton and jet $p_T$'s that are
appropriate for the LHC-13 along with the multiplicities of the same 
(including $b$-$jets$) that
characterize the concerned final states. In addition, these regions also
feature cuts on kinematic variables like the missing transverse energy
($\etmiss$), $\mttwo$ and $H_T$ which are chosen by studying the respective
distributions as shown in figure \ref{fig:kinematic-distribution}. Note that
the plots in figure \ref{fig:kinematic-distribution} are obtained using the
benchmark scenario (BP6) which, as we would find later (see table
\ref{tab:yields}), shows a good sensitivity to the rather characteristic
multi-lepton ($>2$ leptons) final states. Furthermore, the set of kinematic
cuts extracted from these distributions is also found to be optimal for
other benchmark scenarios that we consider in this work.

We find that the LHC Run-I may not be generically sensitive to scenarios having
somewhat low sbottom masses ($\lesssim$ 400 GeV). The benchmark point BP1 having
an R-type sbottom with mass $\approx$ 400 GeV, a near-degenerate pair of
higgsino-like neutralinos ($\ntrltwo$, $\ntrlthree$) with
$m_{\chi_{_{2,3}}}^0 \approx 200$ GeV and with an appreciable
BR[$\sbone \to b \ntrltwothree$] is perfectly allowed by the recent LHC analyses.
This is further confirmed through an explicit check using {\tt CheckMATE}.
Also, $\sbone \equiv \sbleft$ with mass $\approx$ 400 GeV (see scenario BP6)
escapes the current bound that assumes
BR[$\sbone \to b \ntrlone$]=100\%. This is because in the present case, $\sbone$ has
shared branching fractions distributed among available two- and three-body decay
modes, in all of which the final states are different from the one
(2$b$-$jets+\etmiss$) considered by the ATLAS collaboration
\cite{Aad:2015pfx,Khachatryan:2015wza}.
Note that while some such situations discussed in the framework of the MSSM are
perfectly allowed departures from some simplifying assumptions, these are just the
natural expectations in the NMSSM in the presence of a light singlino-like LSP.
%
\begin{table}[!htbp]
\centering
  \begin{tabular}{ ||c|c|c|c|| }
    \hline \hline
 Benchmark  & \multicolumn{3}{c||}{Number of events and signal significance} \\
    \cline{2-4}
    scenario & $\mathrm{SRSSDL}$ & $\mathrm{SR3\ell 2b}$ & 
   $\mathrm{SR4\ell1b}$  \\
\hline \hline
BP1 &         & (88,1200) &  (120, 146) \\ 
    &  $<$ 2  &   2.5     &   9.0       \\ \hline
    
BP2 &          &          &  (56, 146)  \\ 
    &  $<$ 2 &  $<$ 2      &   4.4       \\ \hline
    
BP3 &          &          &  (37, 146)  \\ 
    &  $<$ 2 &  $<$ 2     &   3.0       \\ \hline
    
BP4 & (58, 47) & (158, 1200)  &         \\ 
    &   7.3       & 4.3       &   $<$ 2 \\ \hline
    
BP5 & (36,47) &        &                \\  
    &    4.7  & $<$ 2 &  $<$ 2          \\ \hline

BP6 &         & (240, 1200) &  (80, 146) \\  
    &$<$ 2    &   6.7       &   6.1      \\
\hline \hline
\end{tabular}
\caption{
Number of signal and background events after cuts (given in the parentheses, top
sub-rows) and the corresponding signal significance (the bottom sub-row; for an
integrated luminosity of 300 \fbinv) in different final states for the six
benchmark points at the LHC-13.
} 
\label{tab:yields}
\end{table} 
%

Such NMSSM scenarios, however, can be studied at the LHC-13 in its early phase
in various multi-lepton final states. The signal yields are presented in table
\ref{tab:yields} along with the corresponding expectations for the backgrounds
for an
integrated luminosity of 300 \fbinv. Also indicated in each case is the value
of the signal-significance calculated using the Poisson distribution given by
\beq
\sigma = \sqrt{2\left[(S+B) \, \mathrm{ln}(1+{S \over B})-S \right]}.
\eeq 

To summarize, scenarios with $\sbone \equiv \sbright$ (BP1, BP2 and BP3) are
only sensitive to 4-lepton+$b$-$jet$ (SR4$\ell$1b) final state. Final states
requiring three $b$-$jets$ (e.g., SR3$\ell$3b) are generally less promising
irrespective of the scenarios considered. This is since the corresponding rates
get suppressed by the extra $b$-tag efficiency factors. It should be stressed
here that unravelling such a scenario would require corroborative signatures in
multiple final states. Note that even with 300 \fbinv of data, one could have
not more than two simultaneous final states for which the signal significances 
reach $~5 \sigma$ (SRSSDL and SR3$\ell$2b for scenario BP4 and SR3$\ell$2b and
SR4$\ell$1b for scenario BP6).

In such scenarios with $\tilde{b}_1 \equiv \tilde{b}_R$, the SSDL final state is
not at all sensitive since the decay $\tilde{b}_1 \to t\chi_1^-$ is heavily
suppressed. For an R-type $\sbone$, its decay shares branching among the modes
$\tilde{b}_1 \to b\ntrltwothree$ significantly. For BP1 and BP2, the 4-lepton
final states are reasonably sensitive as $\ntrltwothree$ decays 100\% of the times to
$\ntrlone Z$ and leptons come from the decays of the $Z$-bosons. Still, the
sensitivity is smaller for BP2 sheerly because of a heavier $\sbone$. On the
other hand, scenario BP3, in which decays of $\ntrltwothree$ share branchings
between $Z\ntrlone$ and $h\ntrlone$, naturally loses sensitivity to this final
state.

Benchmark points BP5 and BP6 present the cases where $\sbone \equiv \sbleft$.
For BP5, BR($\tilde{b}_1 \to t\chi_1^-$) is 100\%. Naturally, as discussed
earlier in section \ref{subsec:sbotl}, the most sensitive final state is the
one with SSDL. In BP6, we present a case where 3-body decays of $\tilde{b}_1$
(to $t W^- \ntrlone$ and $b W^+ \charonem$) compete with its 2-body decays
(to $b \ntrlonetwothree$).  This is the reason why BP6, albeit comes with an
L-type $\sbone$, has a poor sensitivity in the SSDL final state. 
On the other hand, it has much better sensitivities in the 3-lepton and the 
4-lepton final states. 

It has been pointed out earlier that a relatively light  L-type sbottom 
will always be accompanied by a (L-type) stop close-by in mass when the mixing 
in the stop sector is not large. In BP5 and BP6 we have such light 
$\stone$-s which are not excluded by the usual stop-search as they share 
branchings among different decay modes like
$t\ntrlonetwothree$ and $b \chi_1^+$ \cite{Aad:2015pfx}. Final states with high
$b$-$jet$ multiplicity like SR3$\ell$3b and SR4$\ell$3b are arising only from
the stop cascade. We find that they are hardly significant with data worth 300
\fbinv.

In scenario BP4, $\sbone$ and $\sbtwo$ are close in mass. As we find, both of
them decay dominantly to $t\chi_1^-$ with branching fractions larger than 90\%.
Note that this is a unique feature in the NMSSM and is in sharp contrast
to the MSSM where, to obtain the mass of the SM-like Higgs boson
in the right ballpark, tan$\beta$ needs to be large and hence $\frac{y_t}{y_b}$
cannot become large enough to make both sbottoms decay dominantly to $t\chi_1^-$.
Naturally, this scenario shows enhanced sensitivity ($\approx 7 \sigma$) to the
SSDL final state as both $\sbone$ and $\sbtwo$ contribute with similar strengths.
This is clearly reflected in the signal corresponding significance as presented in table
\ref{tab:yields}. In addition, BP4 is also found to be moderately sensitive to the 
$3\ell 2b$ final state.

It may be mentioned here that, in general, one may expect contributions to the
final states under consideration from stop pair production as well. However, we
find that except for BP6 these contributions are negligibly small. In BP6,
it is $\sttwo$ that may contribute to 3- and 4-lepton final states but these do not
exceed $\sim 25\%$.

A noteworthy omission in the list of final states is the canonical search mode
involving 2$b$-$jets+\etmiss$. An important upshot of such an NMSSM setup with a
singlino dominated LSP is that a light sbottom, irrespective of its composition,
would not generically have a healthy decay branching fraction to such an LSP.
This would then deplete the event count in the 2$b$-$jet+\etmiss$ final state severely
while multi-lepton final states arising from the cascades of sbottom(s) via
higgsino-like neutralinos would tend to dominate. In contrast to an MSSM
scenario with small $\mu$ ($\sim \mueff$ and $<<M_1, M_2$) thus having an LSP
which is nearly mass-degenerate $\ntrltwo$ (and $\charonemp$), such cascades in
a `natural' NMSSM setup of the present kind could lead to leptons which are hard
enough. As can be expected, the contrast becomes rather prominent when the decay
$\sbone \to t \charonem$ is kinematically forbidden.

\begin{figure}[ht]
\vspace{-0.7in}
\hspace*{-.25in}
\includegraphics[height=0.24\textheight, width=0.58\columnwidth , clip]{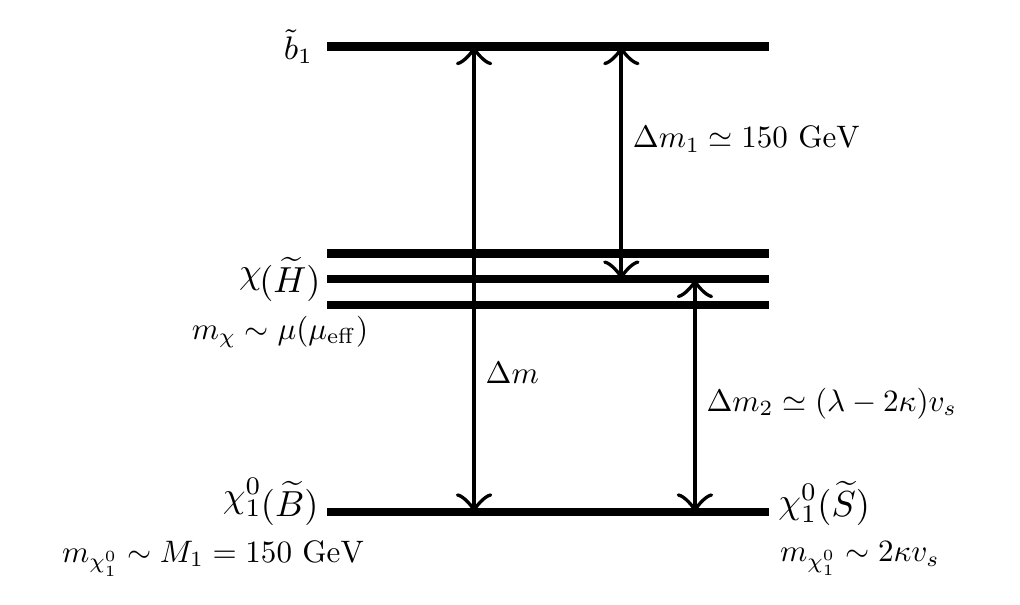}
\hfill
\includegraphics[height=0.30\textheight, width=0.55\columnwidth , clip]{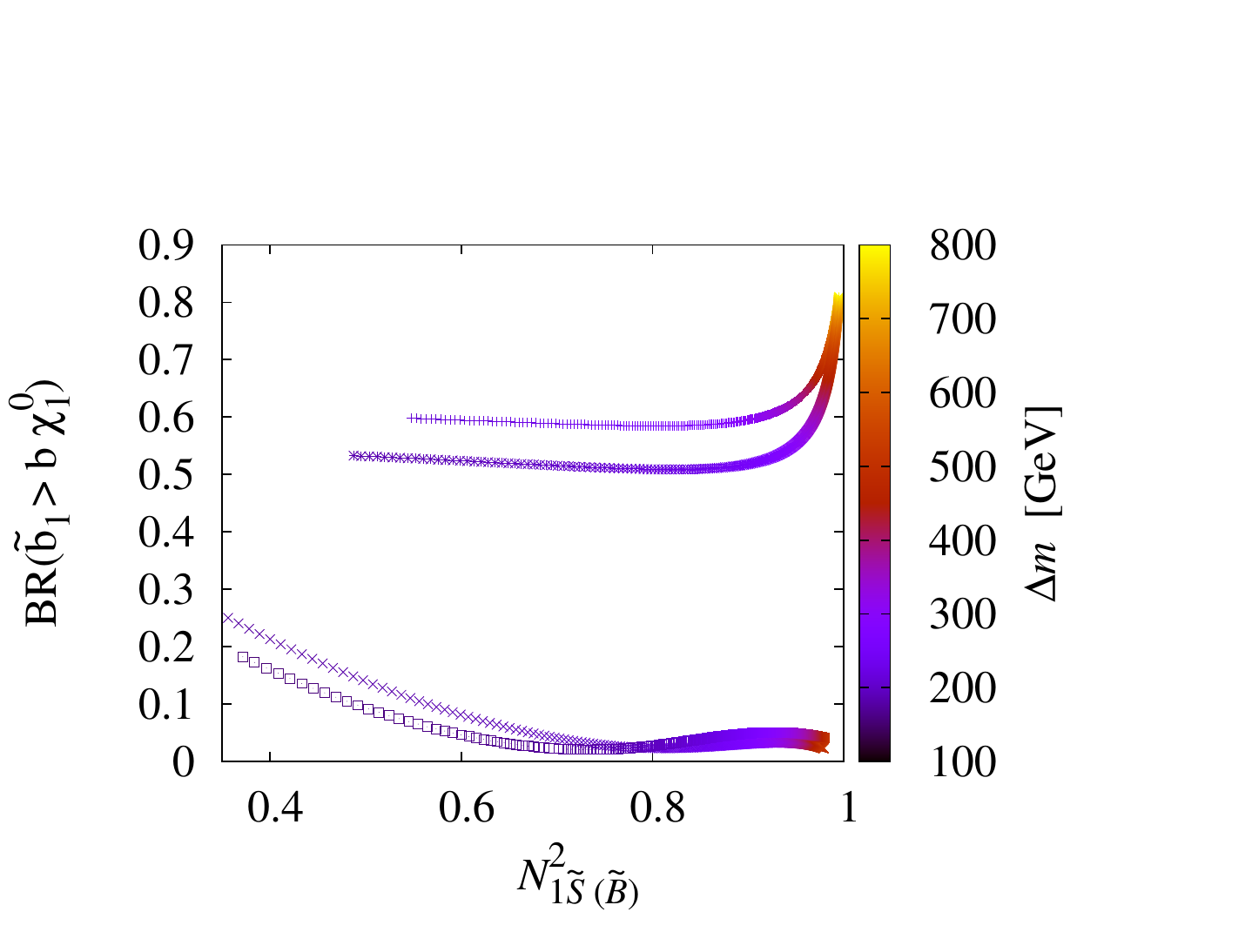}
\vspace*{-0.2in}
\caption{
{\bf Left:} Level diagram representing the spectrum comprising of a relatively
light sbottom and a bino- ($\widetilde{B};$ in the MSSM) or a singlino-like
($\widetilde{S}$; in the NMSSM) LSP along with higgsino-like neutralinos and
chargino sandwiched between them.
{\bf Right:} Variations of the effective branching fraction of a pair of sbottoms
decaying into 2$b$-$jets+\etmiss$ final state as a function of bino (singlino)
content of the LSP, $N_{1\tilde{B}(\tilde{S})}^2$ in the MSSM (NMSSM). The upper
(lower) pair of curves correspond to MSSM (NMSSM). In both cases 
$\mu/\mueff >0 (<0)$ is represented by the upper (lower) curve. The color
palette indicates the simultaneously changing mass-splitting ($\Delta m$) between
$\sbone$ and the LSP. The following fixed values of various parameters are used:
$\lambda=0.7$, $\kappa=0.18$ and $\tan\beta=5$.
}
\label{fig:br-mssm-nmssm}
\end{figure}

In principle, such a depletion in the event rate in the 2$b$-$jets+ \etmiss$
final state, when contrasted against its expected rate in the MSSM, could be
rather illuminating. This is best demonstrated in the variations of the
effective branching fraction of a pair of sbottoms decaying to this final state
in the NMSSM and in the MSSM. The spectral setup adopted for the purpose is
shown in the left of figure \ref{fig:br-mssm-nmssm} in the form of a level
diagram. Note that, in the MSSM case, we deviate from the scheme with
higgsino-like LSP described in the last paragraph and introduce a bino-like LSP
which can have some mixing with the higgsinos. This choice is rather conservative
in the sense that it mimics the NMSSM spectrum thus allowing us to have a
faithful estimate of the extent by which the signals from an analogous MSSM setup
could masquerade as the NMSSM one. As is just pointed out, the mass-splitting
between $\sbone$ and the higgsino-like chargino is restricted to 150 GeV such
that the decay $\sbone \to t \charonem$ is prohibited and $\sbone$ always decays
to bottom quark and neutralinos. In the right of figure \ref{fig:br-mssm-nmssm}
we present the variations of the branching fraction as a function of the
singlino/bino content of the LSP ($N^2_{1 \tilde S/\widetilde B})$ while the
palette shows the simultaneously varying mass-splitting ($\Delta m$) between
$\sbone$ and the LSP.
%

To conform to our broad scenario, we choose $\lambda=0.7$ and $\kappa=0.18$
with $\tan\beta=5$ while $A_{\lambda,\kappa}$ are adjusted to obtain the mass
of the SM-like Higgs boson in the right ballpark. Variations in both $\Delta m$
and $N^2_{1 \tilde S/\widetilde{B}}$ are achieved by varying $|\mueff (\mu)|$
over the range 150-500 GeV. Note that in the NMSSM with $\mueff= \lambda v_S$
and $m_{\tilde{S}} \simeq 2 \kappa v_S$, varying  $\mueff$ results in a varying
$m_{\tilde{S}}$ and hence in a changing LSP mass. At the same time, such a
variation alters the splitting between the higgsino-like states and the LSP
which is roughly given by $\Delta m_2 \approx (\lambda-2\kappa) v_S$,
as indicated in the level diagram of figure \ref{fig:br-mssm-nmssm}.

In the MSSM case we make a conservative choice of $M_1=150$ GeV for which any
chargino mass is allowed \cite{Aad:2015eda}. Further, a minimum difference of
10 GeV between $\mu$ and $M_1$ is required with $|\mu| > M_1$. With this, the
minimum bino fraction in the LSP that we could achieve in the MSSM is
$\sim 50$\% (which is manifested in the abrupt termination of the MSSM curves
on the left side of the right plot of figure \ref{fig:br-mssm-nmssm})
when $|\mu| \simeq 160$ GeV. Note that for a similar higgsino
content of the LSP the bino-higgsino mass difference in the MSSM is usually
smaller than the singlino-higgsino mass difference in the NMSSM. Thus, with
$\msbone = 150 +\mueff (\mu)$, the decay mode $\sbone \to b \ntrlone$ gets
more phase space in the NMSSM when compared to the MSSM case\footnote{Note,
however, that with these choices the lightest sbottom might become as light as
$\sim 300$ GeV which may still be allowed under certain circumstances
\cite{Aad:2015pfx}. In any case, given that the purpose here is to demonstrate
how BR[$\sbone \to b$ LSP] could get heavily suppressed for a singlino-like LSP,
we choose not to get into the intricacies involving the possibility of such a
light sbottom. Furthermore, in this discussion on variation of branching fraction,
the more important parameter is the mass-splitting $\Delta m$ (rather than the
absolute mass of the LSP) and as long as it allows for a large enough $\etmiss$
and render other final state objects visible, the discussion goes through.}.

The right plot of figure \ref{fig:br-mssm-nmssm} shows that the effective
branching fraction to the $2 b + \etmiss$ final state (arising from $\sbone$
pair production) can get severely suppressed in the NMSSM (with light singlino
and higgsinos) when compared to an analogous neutralino spectrum in the MSSM
(with light bino and higgsinos). This is so even though, as pointed out in the
last paragraph, the available phase space is larger in the case of the NMSSM as
compared to the MSSM.
This suppression can clearly be attributed to the absence of any tree-level
coupling of $\sbone$ to a singlino-like LSP in the NMSSM as opposed to the
presence of a tree-level coupling between the sbottom and the bino-like LSP in
the MSSM. Furthermore, it is clear, as expected, that the NMSSM rates are
decreasing with increasing singlino-content of the LSP.
In the MSSM, because of similar couplings of $\sbone \equiv \sbleft$ to bino 
and the higgsinos (for our choice of $\tan\beta$), the branching ratio to the 
LSP is characteristically larger.\footnote{
Further, note that this branching ratio does not initially vary appreciably as 
the bino fraction in the LSP increases since it is accompanied by a similar 
increase in the higgsino component of the NLSP neutralinos which are also 
kinematically accessible to $\sbone \equiv \sbleft$ to decay to. However, with 
increasing $\mu$, in our setup, the mass-splitting between $\sbleft$ and the 
LSP grows while the same with respect to the NLSPs remains roughly unaltered. 
Thus, the growing phase space explains the quick pick-up in 
BR[$\sbone \to b$ LSP] at the right edge of the plot.}

%

Note that only L-type $\sbone$ has been considered in figure \ref{fig:br-mssm-nmssm}.
R-type $\sbone$ with a larger hypercharge would always enhance the partial width
for $\sbone \to b \ntrlone$ in the MSSM with a bino-like LSP thus making the
difference with NMSSM more drastic. Similarly, our generic choice of a smaller
$\tan\beta$ would also lead to a smaller $y_b$ thus suppressing the $\sbone$
decays to higgsino-like states in the MSSM which would further favor its decay to
the bino-like LSP in such a scenario.
%

\section{Conclusions}
\label{sec:conclusions}
In this work we discuss the characteristic
phenomenology that the sbottoms could
derive in a `natural' NMSSM setup even though they do not actively address the
issue of naturalness. We point out the generic possibility of having light
sbottom(s) in such a scenario which could evade current
bounds from the LHC experiments. The NMSSM setup considered in this work is
characterized by relatively small $\mueff$ and light stop(s) which facilitate
compliance with the standard notion of naturalness. The latter requires one to
consider moderate to large values of $\lambda$ in order to have the mass of the
SM-like Higgs boson in the experimentally observed range.

It is stressed how the two ingredients that render the setup `natural', i.e.,
small $\mueff$ ($|\mueff| \lesssim 350$ GeV) and large $\lambda$ ($\sim 0.6-0.7$),
could have an immediate impact on the sector comprising of lighter neutralinos.
While this combination tends to require a small $v_S$, large $\lambda$ by itself
implies small $\kappa$ once absence of Landau pole below a high (unification)
scale is demanded. These, in turn, drive the singlino mass ($= 2 \kappa v_S$) small
thus making it comparable or even smaller than $\mueff$. Such a situation, in the
presence of a large $\lambda$, could potentially make the lighter neutralino
sector phenomenologically nontrivial.  As far as the LSP is concerned, its
composition could range between that like a nearly-pure singlino to a heavy mixture
of singlino and higgsinos. This is accompanied by two neutralinos (immediately
heavier than the LSP) whose compositions range between pure higgsinos and a similar
mixture of higgsinos and singlino as in the case of the LSP. In the present
context, this could crucially alter the ways in which sbottoms could decay.
In our `simplified' scenario we assume the soft gaugino masses, $M_1$ and $M_2$
(and $M_3$ as well) to be rather heavy. The lighter chargino is thus higgsino-like 
and the heavier electroweakinos and the gluino all decouple.

A further interesting observation in such an NMSSM framework (with large
$\lambda$ and light stops) pertains to the opening up of the rather low $\tan\beta$
($\lesssim 2$) regime which is disfavored in the MSSM as it fails to find the
mass of the SM-like Higgs boson in the right ballpark. A $\tan\beta$ value as
low as 2 boosts the ratio $y_t \over y_b$ thus quickly enhancing the $y_t$
driven decays of $\sbone$ over the $y_b$ induced ones. This is found to be an
important issue in the phenomenology of an L-type sbottom whose decay to
higgsino-like chargino and a top quark is induced by $y_t$. Even more
interestingly, the effect can be so strong that a 3-body decay of an L-type
sbottom involving a $y_t$-driven vertex could overwhelm its $y_b$-induced
2-body decays to $b \ntrlonetwothree$ for low to moderate virtuality of the
propagating state ($\charonemp$ or the top quark). On the other hand, an
enhanced $y_t \over y_b$ could also ensure that a mostly R-type sbottom, with a
tiny admixture of L-component, can have a significant branching fraction to
$t \charonem$.

We choose a few benchmark NMSSM scenarios satisfying the criteria mentioned
above.  These have the singlino-content of the LSP varying between 75\% and
90\%. The lighter sbottom can be (almost) purely R- or L-type. As pointed
out in the last paragraph, we also discuss the interesting case of R-dominated
lighter sbottom with a tiny admixture of the L-component. The mass of the 
lighter sbottom lies between 400 GeV and 600 GeV in these representative scenarios.

On the other hand, the heavier sbottom is arranged to be rather heavy
($\gtrsim 1.5$ TeV), in general. However, we showed (scenario BP4) that
requiring even a small admixture of the L-component in an otherwise R-dominated
light sbottom would result in a heavier sbottom not so different in mass from
its lighter partner.
Furthermore, scenarios with an L-type light sbottom inevitably have a stop
squark in the spectrum of similar mass and hence should be simultaneously
accessible at the LHC. In these two cases, a given final state may receive
some limited contributions from more than one excitation (the heavier
sbottom and the stop squarks) thus making its interpretation potentially
involved.  In the latter case, however, it is important to ensure that such
a light sbottom indeed escapes the indirect bound that could be derived from
non-observation of an accompanying light stop. Evading such a bound could be
possible when the stop sector has an appreciable mixing and/or, for an unmixed
L-type $\stone$, if it could decay to heavier neutralinos. The former would,
in turn, prompt the lighter stop to decay in multiple ways, with shared
branching fractions.

For an L-type lighter sbottom, its $t\charonem$ decay mode, if kinematically
open, would dominate and would result in a healthy SSDL rate. A much enhanced
SSDL rate might point to both sbottoms contributing to this final state with
one of the sbottoms being L-type, for all practical purposes, while the mostly
R-type one has a slight admixture of $\sbleft$. If $t\charonem$ decay mode is
not kinematically accessible, an L-type lighter sbottom would rather show up
strongly in several multi-lepton final states. For a light NMSSM sbottom which
is mostly R-type, the only resort to see its signature, in such a scenario,
could be in the four-lepton final state.

The bottom line of the present study is that the signals of sbottoms in a natural
NMSSM framework could be characteristically stubborn in not showing up promptly
at the LHC experiments. This feature is crucially governed by the nature of the
LSP. At the LHC Run-I this might have helped sbottoms escape the searches. 
On the other hand, the LHC-13 would require moderately large data (worth $\sim$300
\fbinv) to hint/establish their presence in various multi-lepton plus $b$-$jets$
final states with $\etmiss$. There, it would be further corroborative if one
experiences a dearth of events in the 2$b$-$jets+\etmiss$. Improved techniques
proposed recently \cite{Schlaffer:2016axo} could help sharpen the search for the
sbottoms (and stops) in an involved situation. These observations should be
helpful in planning future experimental analyses to uncover possible spectral
configurations with light sbottoms and an LSP with a singlino-admixture in a
`natural' NMSSM setup.
%
\acknowledgments
JB and AC are partially supported by funding available from the Department of
Atomic Energy,  Government of India for the Regional Centre for Accelerator-based
Particle Physics (RECAPP), Harish-Chandra Research Institute. AC acknowledges
financial support from the Department of Science and Technology, Government of
India through the INSPIRE Faculty Award /2016/DST/INSPIRE/04/2015/000110. The
authors acknowledge the use of the cluster computing setup available at RECAPP
and at the High Performance Computing facility of HRI. JB like to thank B. Fuks
for very helpful discussions. The authors thank U. Chattopadhyay for his kind
hospitality at the Theoretical Physics Group of the Indian Association for the
Cultivation of Science, Kolkata where a part of the work is done. They also thank
Amit Khulve and Ravindra Yadav for technical support.
%

\bibliographystyle{JHEP} 
\bibliography{sbot-jhep.bbl}
\end{document}